\documentstyle[pre,aps,floats]{revtex}
\input epsfig.sty
\begin{document}
\preprint{P-98-08-**}
\draft
\title{Topological defects and the short-distance behavior of the 
structure factor \hfil\break 
in nematic liquid crystals}
\author{Martin Zapotocky\cite{MZ} }
\address{
Department of Physics and Astronomy,\\
University of Pennsylvania, Philadelphia, Pennsylvania 19104}
\author{Paul M.~Goldbart\cite{PMG} }
\address{
Department of Physics and Materials Research Laboratory,\\
University of Illinois at Urbana-Champaign, Urbana, Illinois 61801}
\date{\today}
\maketitle
\begin{abstract}

The scattering of light at large wave-vector magnitudes $k$ in nematic 
systems containing topological defects is investigated theoretically. 
At large $k$ the structure factor $S(k)$ is dominated by power-law
contributions originating from singular order-parameter variations
associated with topological defects and from transverse thermal
fluctuations of the nematic director. 
 These defects (nematic
disclinations and hedgehogs) lead to contributions of the form $\rho A
k^{-\xi}$ (``the Porod tail''), where $\rho$ is the number density of a
given type of defect, $A$ is a dimensionless Porod amplitude, and $\xi$ is an
integer-valued Porod exponent. The Porod amplitudes and
exponents are calculated 
for all types of topologically stable defects occurring in
uniaxial and biaxial nematics in two or three spatial dimensions.  
The range of wave-vectors in which the contributions to the
scattering intensity due to defects dominate the contribution due to
thermal fluctuations is estimated, and it is concluded
 that for experimentally accessible
defect densities the range of observability of the Porod tail extends
over one to three decades in scattering wave-vector magnitude $k$.  
Available experimental
results on phase ordering in uniaxial nematics are analyzed, and 
applications of our results are suggested for light-scattering studies of other
nematic systems containing numerous defects.

\end{abstract}
%

\section{Introduction}
\label{SEC:Introduction}

When a liquid crystalline material is brought
from the isotropic phase to the nematic phase, the turbidity (i.e.,
the total intensity of scattered light) of the material increases by a
factor of the order of $10^6$, 
and the sample appears cloudy~\cite{de Gennes}.  
In samples in which the nematic ordering is well aligned, 
this intense scattering is due predominantly to
thermal fluctuations in the orientation of the nematic director, and 
has been extensively studied and reported in the literature
\cite{de Gennes,Chatelain48,Gennes68,Langevin75}.  
If, on the other hand, the  sample is not well aligned and contains a 
large number of topological
defects (disclinations, nematic hedgehogs, or surface defects),
the turbidity becomes dominated by scattering from essentially static 
director 
inhomogeneities associated with these defects \cite{Langevin75}.

A particularly visible manifestation of defect-associated 
scattering is the so-called ``Porod tail'' contribution to 
the scattering intensity. For sufficiently large magnitudes $k$ of the 
scattering wave-vector ${\bf k}$, the scattered light intensity $I({\bf k})$  
of a nematic system containing topological defects decays as 
a power law $k^{-\chi}$, where $\chi$ is an integer-valued  exponent 
that depends on the dominant type of defect present in the system. 
Such behavior was confirmed and the values of $\chi$ have been 
discussed in both the experimental \cite{Wong92,Wong93} and 
the theoretical \cite{Bray-nematic} literature on the kinetics of phase 
ordering of nematic systems containing numerous defects 
generated during a quench from the isotropic to the nematic 
phase. A full calculation of the form of the power-law tails 
associated with nematic defects, and an evaluation of the 
relative importance of these contributions compared to the $k^{-2}$ 
contribution associated with thermal fluctuations have, 
however, not yet been given in the literature, 
and form the subject matter of the present Paper.

Our results are, in principle, applicable to any nematic system
containing topological defects, but should be especially useful in
interpreting detailed measurements of scattered-light intensity in
highly disordered systems (i.e., those containing numerous
defects). Such configurations arise in low-molecular-weight nematics
for example when the transition from the isotropic to the nematic
phase is sudden (e.g., induced by a temperature quench)
\cite{Chuang91,Chuang93,Wong92,Wong93,Orihara93,Bowick94}, 
or when an originally well-aligned nematic sample is put in a
sufficiently strong shear flow \cite{Larson96} or under high
alternating voltage \cite{Kai90}.  In polymer nematics, numerous
defects often persist even in the absence of any external agents
\cite{Graziano84,Kleman83}, and significantly affect the mechanical
\cite{Larson_book} and electro-optical \cite{Blinov94} properties of
these materials.  Our results permit the extraction of information on
the type and number of topological defects present in such systems and
the monitoring of the dynamics of the defects.  As our results are
exact in the appropriate scattering wave-vector range, they can also
be used to test the validity of analytical theories and simulations of
phase ordering in nematics.

This Paper is organized as follows. In Sec.~\ref{SEC:GenSk} we
discuss the general issue of the origins of power-law contributions to the
structure factor $S(k)$ (i.e. the Fourier-transformed order-parameter
correlation function) in systems possessing continuous symmetries. In general,
two types of power-law contributions of quite distinct origins are
present: those due to transverse thermal fluctuations of the order
parameter, and those due to topological defects. The
order-parameter variations associated with topological defects give
rise, at sufficiently large $k$, to contributions of the form $\rho A k^{-\xi}$
(the Porod tail), where $\rho$ is the number density of a given
type of defect, $A$ is a dimensionless Porod amplitude and $\xi$ is an
integer-valued Porod exponent. In Sec.~\ref{SEC:OnSymm} we re-derive and
generalize some of the results of Bray and Humayun~\cite{Bray-amplitudes} 
for Porod amplitudes and exponents in the
$O(N)$ vector model by using a somewhat different and computationally
simpler method.  In the remainder of the Paper, we use this method to
calculate Porod amplitudes and exponents for topological defects in
uniaxial and biaxial nematic liquid crystals.  The forms of the Porod tail 
that correspond to hedgehog defects, disclination lines, and ring
defects in uniaxial nematics are calculated in subsections
\ref{SEC:hedge}--\ref{SEC:RingsInUni}.  A case of special interest is that of a
wedge-type ring defect (i.e. disclination loop) in a uniaxial nematic,
where two separate Porod regimes (having distinct exponents and
amplitudes) arise for length scales larger than and smaller than the
radius of the disclination loop. The presence of two Porod regimes is
specific to the nematic case, and does not occur for any type of
defect in $O(N)$ vector model systems.  In Sec.~\ref{SEC:thermal} we
discuss the influence of transverse {\it thermal\/} fluctuations of
the nematic director on the large-$k$ behavior of the structure factor.
In Sec.~\ref{SEC:exper} we use the results of
Secs.~\ref{SEC:intronem}--\ref{SEC:thermal} to analyze the results of
the light-scattering experiments of Refs.~\cite{Wong92,Wong93},
which investigate the process of phase-ordering kinetics in uniaxial
nematics.  In Sec.~\ref{SEC:BiaxNemSys} we generalize our results
for the uniaxial nematic to the case of non-abelian defects in {\it
biaxial} nematics, the dynamics of which have recently been
investigated experimentally~\cite{De'Neve92} and theoretically~\cite{2dnem}.  
A general discussion of corrections to 
the results derived in this Paper that may arise from effects such as
defect interactions, defect curvature, and the presence of the defect
core, is given in Sec.~\ref{SEC:CorrToPorod}.  We conclude, in
Sec.~\ref{SEC:Concl}, with a summary of our results and suggestions
for their use in the analysis of experimental data.


\section{Generalized Porod law}
\label{SEC:GenSk}

Consider an ordered system characterized by an order-parameter
field ${\bf\Phi}({\bf r})$, 
where ${\bf\Phi}$ is an $N$-component vector in order parameter space 
and ${\bf r}$ is the radius-vector in $d$-dimensional real space.   
The principal quantity of interest in many theoretical and 
experimental investigations of ordered systems is the structure factor
$S({\bf k})$, which is the Fourier transform of the real-space
correlation function $C({\bf r})$ of the order parameter. 
This quantity, which is directly measurable via the appropriate
scattering experiment, characterizes the degree of inhomogeneity of the
order parameter at length scales of order $k^{-1}$.
By definition, 
\begin{equation}
S({\bf k}) \equiv
\int d^{d}r\,e^{i{\bf k}\cdot{\bf r}}\,C({\bf r})\,,
\end{equation}
with the real-space correlation function $C({\bf r})$ given by
\begin{equation}
C({\bf r}) = {1 \over M_{O(N)}} 
\int d^dx\, {\bf \Phi}({\bf x})\cdot 
{\bf \Phi}({\bf x}+{\bf r}) \,,
\label{EQ:defcr}
\end{equation}
where the integration is taken over the whole system.
Here $M_{\rm O(N)}$ is a normalization factor,
\begin{equation}
M_{\rm O(N)} \equiv \int d^dx\,{\bf \Phi}({\bf x}) \cdot 
{\bf \Phi}({\bf x}) \,,
\label{EQ:norm}
\end{equation}
chosen to ensure that $C({\bf r})\vert_{{\bf r}={\bf 0}}=1$.
Notice that with this choice of normalization $C({\bf r})$ is dimensionless,
whereas $S({\bf k})$ has the dimensions of a volume.
Equivalently \cite{FT}, the structure factor may be expressed as
\begin{equation}
S({\bf k}) = {1 \over M_{O(N)}} {\bf \Phi}({\bf k}) \cdot 
{\bf \Phi}({\bf -k}) \,, 
\label{EQ:wigner}
\end{equation}
where $\Phi({\bf k})$ is the Fourier-transformed order parameter,
\begin{equation}
{\bf\Phi}({\bf k}) \equiv \int d^dr\,
e^{i {\bf k}\cdot{\bf r}}
{\bf\Phi}({\bf r}) \,. 
\label{EQ:ftphi}
\end{equation}
For systems that are isotropic on macroscopic scales (such as a bulk
system undergoing phase ordering), $S({\bf k})$ depends on ${\bf k}$
through the magnitude $k=|{\bf k}|$ only. 

We now discuss the contributions 
to $S({\bf k})$ that have the form of a power law ( i.e.,  
$a k^{-\chi}$) where $a$ and $\chi$ are $k$-independent constants. 
First, we briefly recall the effect of thermal fluctuations. 
 Consider
an $O(N)$ vector model system (with $N \geq 2$) in the ordered phase with the
thermally-averaged value  of the order
parameter ${\bf \Phi}({\bf x})$ 
pointing in the $x_1$ direction. Due to the $O(N)$ symmetry,
fluctuations in ${\Phi}_i({\bf x})$ perpendicular to the $x_1$
direction (i.e., transverse fluctuations) 
cost an arbitrarily low amount of free energy for any large-length-scale 
fluctuations; this leads to 
strong scattering at small $k$ in the whole 
temperature range of the ordered phase. 
[In contrast, strong low-$k$ scattering in systems having 
a {\it scalar\/} order 
parameter (for which transverse fluctuations do not exist) 
arises only in the immediate 
vicinity of a critical point (``critical opalescence'') and is due 
to longitudinal fluctuations 
(i.e., fluctuations in the order parameter {\it magnitude\/}).]
For our purposes, the important property of transverse fluctuations 
is that they occur on {\it all\/} length scales between the 
size of the system (small $k$) and the microscopic coherence length of 
the order parameter (large $k$). 
In a system of spatial dimension $d > 2$ \cite{twodimthermal} and 
described by the standard form of the gradient free energy~$E$
\begin{equation}
E = {\kappa \over 2} \int d^3\,x \, \nabla_i \Phi_j({\bf x}) 
\nabla_i \Phi_j({\bf x})\,
\label{ONenergy}
\end{equation}
(where summation over the indices i and j is implied),
the contribution to the structure factor due to 
thermal fluctuations has the form $a(T)
k^{-2}$ with a temperature-dependent amplitude given by the 
microscopic length scale $a(T)=k_B T/\kappa$ \cite{Nigel_book}. 

Next, we discuss power-law contributions to the structure factor 
that arise from topological defects. 
Topologically stable defects occur in all $O(N)$ vector model systems 
whenever $d \ge N$.
The order-parameter variations associated with the defects give
rise to the Porod-tail form of the structure factor, 
$S(k) \sim k^{-\chi}$, for sufficiently large values of $k$. 
In contrast to the 
thermal-fluctuation contribution, the power-law contributions to $S(k)$ 
due to topological 
defects do not depend on 
temperature (provided the defect structure is not affected by the temperature)
and occur in systems with discrete symmetries as well as in 
those with continuous symmetries.

In scalar systems, this power-law behavior has long been understood as
arising from abrupt changes in the order parameter across domain walls
present in the system~\cite{Debye57,Porod82}.  Consider, for simplicity,
a one-dimensional system in which the order parameter suffers a kink.  
Viewed on length scales larger
than the kink width (which is given by the coherence length of the order
parameter), the order parameter abruptly changes from $-1$ to $+1$ at the
kink location.  Consequently, the integrand in Eq.~(\ref{EQ:defcr}) 
equals $-$1 within the distance $|r|$ from the kink (domain wall)
location and $+1$ elsewhere in the system.  This results in the
normalized real-space correlation function of the form $C(r) = 1 - 2
|r|$. Notice that $C(r)$ is non-analytic at $r=0$, reflecting the
singular variation of the order parameter at the kink.  The
non-analytic part of $C(r)$, after Fourier-transforming, yields a
power-law form of the structure factor, $S(k) \sim k^{-1}$, valid
for values of $k$ larger than the inverse system size and smaller than
the inverse kink width.

For domain walls in dimensions $d$ higher then one, additional factors
of $1/k$ arise due to the spatial extent of the domain walls, with the
result (averaged over domain-wall orientations) $S(k)\propto
k^{-(d+1)}$. In regions between the domain walls, the order parameter
does not vary, and therefore there are no extra contributions to the
structure factor for $k$ values larger than the inverse domain
size. We thus obtain a structure factor that has a power-law form for
$k$ in the range between the inverse average domain wall separation
and the inverse domain wall width, with the pre-factor proportional to
the total domain wall area in the system.

Bray and Humayun~\cite{Bray-amplitudes} investigated in detail how the
singular variation of the order parameter around a point or line defect
in a system with {\it continuous} symmetry will likewise lead to an asymptotic
power-law dependence on $k$ of the structure factor $S(k)$.  
In Ref.~\cite{Bray-amplitudes}, 
they calculated exactly the exponent and the amplitude in the
contribution $A k^{- \chi}$ to the structure factor coming from an
isolated defect order-parameter
configuration in the $O(N)$ vector model.  They also argued that for 
sufficiently large $k$, the
structure factor of a system containing a number of topological defects
can be obtained by multiplying $A k^{-\chi}$ by the number density of defects. 
The reason is as follows.  Variations in the order parameter on very
short scales [probed by $S(k)$ with $k$ large] occur only in the
singular regions surrounding the defect cores, which are separated 
by regions that barely contribute to $S(k)$. 
It is therefore
appropriate to treat the contribution from each defect independently,
and as coming from an isolated defect, 
{\it provided that the configuration in the vicinity of
the defect core is not affected by the presence of other defects\/}.
While this assumption is known to be valid 
in some special cases [such as the $O(2)$ vector model], it is in general 
non-trivial. For example, it was
argued in Ref.~\cite{Ostlund} that, for the case of point defects in the
$O(3)$ vector model in $d=3$, a highly asymmetrical, string-like configuration
develops in between the defects. Nevertheless, 
the prediction for $S(k)$ based on the
assumption of undeformed defect configurations has been found to be in good
agreement with numerical simulations in several $O(N)$ vector model 
phase-ordering 
systems~\cite{Blundell94}.  In the calculations we present in
Secs.~\ref{SEC:OnSymm}--\ref{SEC:BiaxNemSys}, we shall assume that 
the regions close to the defect
cores in a phase ordering system {\it do\/} 
possess the structure of an isolated
defect, and in Sec.~\ref{SEC:CorrToPorod} 
we shall comment on the nature of possible corrections to
our results.

The discussion in the preceding paragraphs leads us to the following
formulation of the {\it generalized Porod law\/} for the orientation-averaged 
structure factor
at zero temperature of a system containing 
topological defects:
\begin{equation} 
S(k)\sim A \rho k^{-\chi}\,,
\label{genporod}
\end{equation}
where $\rho$ is the {\it defect density\/} (i.e., the domain wall
area, string length, or number of point defects, per unit volume of
the system), and $A$ is a dimensionless amplitude characterizing the
given type of defect.  The generalized Porod law is expected to be
valid in the range $L^{-1}\ll k\ll\xi^{-1}$, where $L$ is the average
separation between defects present in the system and $\xi$ is the size
of the defect cores. Specific cases exemplifying the validity of
Eq.~(\ref{genporod}) will be given throughout the rest of the paper.
The value of the Porod exponent $\chi$ can be anticipated on purely 
dimensional grounds. Denoting by $d$ the spatial 
dimensionality of the system and by $D$
the dimensionality of the topological defect (that is, $D=0$ for
point defects, $D=1$ for string defects, and $D=2$ for domain walls), 
we see that the defect density $\rho$ in
Eq.~(\ref{genporod}) scales as $L^D/L^d$, whereas the structure factor
$S(k)$ has the dimension of $L^d$. Consequently, the exponent $\chi$ 
is given simply by $2 d - D$ (see Sec.~\ref{SEC:CorrToPorod} for a more 
precise derivation of this result). 
  In contrast, the dimensionless
amplitude $A$ depends in more detail on the order parameter in
question. For example, we shall see that the amplitudes $A$ for the
nematic hedgehog and for the $O(3)$ monopole in $d=3$ differ
substantially.

Nematic liquid crystals represent an interesting case where several
distinct power-law contributions to the structure factor can be
present simultaneously in the same system.  These contributions are
discussed in detail in Secs.~\ref{SEC:hedge}--\ref{SEC:thermal} and
are due to line defects (disclinations), 
point defects (nematic hedgehogs) and thermal fluctuations. 
The competition among these terms can
lead to a variety of forms of the tail of the structure factor appropriate to
distinct physical conditions. Before describing in detail the case of
nematic liquid crystals, however, we illustrate our computational
techniques on the case of Porod amplitudes and exponents in
$O(N)$-symmetric vector-model systems.

\section{$O(N)$ symmetric vector systems}
\label{SEC:OnSymm}

The Porod exponent $\chi$ and amplitude $A$ of a contribution to 
$S({\bf k})$ from a single defect in the configuration of an $O(N)$ vector
model was computed in Ref.~\cite{Bray-amplitudes} by first calculating
the real-space correlation function $C({\bf r})$, Eq.~(\ref{EQ:defcr}),
and then Fourier-transforming the result to obtain $S({\bf k})$.
During this calculation, certain analytic terms (see
Ref.~\cite{Bray-amplitudes}), which do not contribute to the power-law
part of the Fourier transform, were discarded;
 also, for $O(N)$ systems with $N$ even,
the calculation was complicated by the appearance of logarithmic terms in
the intermediate result for $C({\bf r})$.  We shall show below that
the results of Ref.~\cite{Bray-amplitudes} can be obtained in a more
simple way by working directly in Fourier space, i.e., by first
Fourier-transforming the order-parameter configuration, and then using
Eq.~(\ref{EQ:wigner}) to obtain the structure factor \cite{Nigel}. Besides
being shorter, this method does not discard any terms, and does not
involve any logarithmic terms.

\subsection{Point defects in O(N) symmetric vector systems}
\label{SEC:PDinOnSymm}

We first perform the calculation for the case of the $O(N)$ vector model
system in $d=N$ spatial dimensions. In this case, only point defects
(i.e., monopoles) are topologically stable; moreover, for $d=N=2$, 
monopoles with charge 
amplitude $|Q|>1$ are energetically unstable towards the decay into 
monopoles with charge $+1$ or $-1$. In the following, we consider the case 
$Q=+1$; an identical result for the structure factor is obtained in the 
case $Q=-1$. 

The configuration of the $Q=1$ (radial) monopole in a system with the free 
energy given by Eq.~(\ref{ONenergy}) is described by 
\begin{equation}
{\bf\Phi}({\bf r}) = {{\bf r} \over r}\,.
\label{point}
\end{equation}
Note that we have omitted any spatial variation in the amplitude 
$|{\bf\Phi}|$ of 
the order parameter.  This corresponds to our focusing on length scales 
larger that the so-called core of the defect (i.e., the region of space 
in which the competition 
between condensation and gradient free energies gives rise to 
significant modification of $|{\bf\Phi}|$). For a configuration having  
constant $|{\bf\Phi}|$, the correlation function Eq.~(\ref{EQ:defcr}) 
and the structure factor Eq.~(\ref{EQ:wigner}) do not depend 
on $|{\bf\Phi}|$; consequently, we have chosen 
$|{\bf\Phi}|=1$ for our calculation. 

Due to spherical symmetry, the Fourier transform of ${\bf \Phi}({\bf r})$ 
takes the form
\begin{equation}
{\bf\Phi}({\bf k}) = {\bf k}\,f(k^{2})\,,
\label{pointsk}
\end{equation}
where $f$ is a certain scalar function of the scalar $k^{2}$. 
By taking the scalar product of Eq.~(\ref{pointsk}) with ${\bf k}$ and 
solving for $f$ we see that we may write 
\begin{equation}
{\bf\Phi}({\bf k})
={\bf k}\,k^{-2}\,
\int d^{d}r\,r^{-1}\,
 {\bf k}\cdot{\bf r}\,{\rm e}^{i{\bf k}\cdot{\bf r}}
=-i{\bf k}\,k^{-2}
 {\partial\over{\partial\lambda}}\bigg\vert_{\lambda=1}
 F(\lambda^{2}k^{2})\,,
\label{EQ:WithF}
\end{equation}
where the function $F$ is defined via
\begin{equation}
F(\lambda^{2}k^{2})\equiv
\int d^{d}r\,r^{-1}\,
{\rm e}^{i\lambda{\bf k}\cdot{\bf r}}\,, 
\end{equation}
i.e., $F$ is the $d$-dimensional Fourier transform of the 
Coulomb potential $r^{-1}$, which can readily be shown \cite{FT}, 
for ${\bf k}\ne{\bf 0}$, to be 
\begin{equation}
F(\lambda^{2}k^{2})=
{(4\pi)^{(d-1)/2}
\over{k^{d-1}}}
\,\Gamma\left({d-1\over{2}}\right)\,,
\end{equation}
where $\Gamma(x)$ is the standard Gamma-function.
By evaluating the derivative of $F$, and inserting it into 
Eq.~(\ref{EQ:WithF}), we find  
\begin{equation}
{\bf\Phi}({\bf k})
=i(d-1)(4\pi)^{(d-1)/2}
 \,\Gamma\left({d-1\over{2}}\right)
 {{\bf k}\over{k^{d+1}}}.
\end{equation} 
The normalization factor $M_{\rm O(N)}$, given by Eq.~(\ref{EQ:norm}), 
is the volume $V$
of the system.  It follows from Eq.~(\ref{EQ:wigner}) 
that the structure factor $S(k)$ of
a single point defect is then given by
\begin{equation}
S(k)
={1\over{V}}{1\over{\pi}}(4\pi)^{d}
 \,\Gamma^{2}\left({d+1\over{2}}\right)
 {1\over{k^{2d}}}\,.
\label{EQ:MagRes}
\end{equation}
Specifically, we obtain $S(k) = V^{-1} 4 \pi^2 k^{-4}$ for an $O(2)$ 
vortex in $d=2$, 
and $S(k) = V^{-1} 12 \pi^3 k^{-4}$ for an $O(3)$ monopole in $d=3$. 
Note that the result 
Eq.~(\ref{EQ:MagRes}) does not depend on the magnitude $|{\bf\Phi}|$ 
of the order parameter 
(we have chosen $|{\bf\Phi}|=1$ in this subsection)
as both the normalization factor Eq.~(\ref{EQ:norm})
and the numerator in Eq.~(\ref{EQ:wigner}) 
are proportional to $|{\bf\Phi}|^2$. 

For a system having $\rho$ defects per unit volume, the normalization factor 
factor $V^{-1}$ is 
replaced by $\rho$.  The expression Eq.~(\ref{EQ:MagRes}) is then 
in agreement with the result for $S(k)$ given in Ref.~\cite{Bray-amplitudes}. 
It should be noted that by fully exploiting the spherical symmetry at hand, 
the method used above enables us to pass from a calculation involving
vector quantities to a calculation involving only scalars. In 
Sec.~\ref{SEC:hedge}, we shall similarly be able to pass from a tensorial 
calculation, appropriate for the case of the nematic order parameter,
to a scalar one.

\subsection{Vortex lines in three-dimensional O(2) vector systems}
\label{SEC:VortexLines}

We now turn to the case of line defects in $O(N)$ vector systems, 
specifically concentrating on the physically prominent case of 
vortex lines in three-dimensional $O(2)$ vector systems. 

In the case of the spherically-symmetric point defect in the $O(N)$
vector system, investigated in the previous subsection, the final
result, Eq.~(\ref{EQ:MagRes}), is independent of the orientation of the
scattering 
wave-vector ${\bf k}$, due to the $O(N)$ symmetry. In contrast, the
structure factor of a segment of a line defect depends on the angle
between ${\bf k}$ and the orientation of the line segment; 
specifically, $S({\bf k})$
is dominated by contributions from defect segments that are
perpendicular to the vector ${\bf k}$, as there are no short-distance
order-parameter variations in the direction along the defect line.  

We first discuss the structure factor of a single straight vortex line.
Consider a vortex-line segment in $d=3$ having its core
located on the line $(x,y)=(0,0)$, and extending from $z=-L/2$ to $z=L/2$. 
The order parameter ${\bf\Phi}$ does not depend on $z$, and the 
minimum-energy 
configuration of ${\bf\Phi}$ in the $xy$ plane is identical to the 
configuration of a point defect in the corresponding $O(2)$ vector 
model in $d=2$. Thus, the structure factor is given by
\begin{equation}
S({\bf k})
={\bf\Phi}({\bf k})\cdot{\bf\Phi}(-{\bf k})
=S^{(2)}(k_x,k_y)
\int_{-L/2}^{L/2}dz \,e^{ ik_z z } 
\int_{-L/2}^{L/2}dz'\,e^{-ik_z z'}\,,
\label{stringsk}
\end{equation}
where $S^{(2)}(k_x,k_y)$ is the structure factor for the $d=2$ point
defect configuration. From Porod's law in $d=2$ we have 
$S^{(2)}(k_x,k_y)
=A^{(2)}/[V^{(2)} (k_x^2+k_y^2)^{2}]$, 
where $A^{(2)}$ is a dimensionless constant, 
$V^{(2)}$ is the system area, 
and $\theta$ is the angle between ${\bf k}$ and the orientation of the 
defect line. 
We now let $L\rightarrow\infty$ and calculate the structure factor per
unit length of the defect, $S_{\rm seg}(k)=\lim_{L\to\infty} S({\bf k})/L$. 
By using the identity 
\begin{equation}
\lim_{L\to\infty}L^{-1}
\int_{-L/2}^{L/2}dz\,\exp( iuz )
\int_{-L/2}^{L/2}dz' \exp(-iuz')
=2\pi\delta(u) \,,
\end{equation}
we immediately obtain 
\begin{equation}
S_{\rm seg}({\bf k})={2 \pi A^{(2)}\over V}{1\over k^4 \sin^4{\theta}} \,
\delta(k \cos{\theta})\,.
\label{string}
\end{equation}

Eq.~(\ref{string}) can be used as the basis for the evaluation 
of the structure 
factor of an arbitrary vortex-loop configuration, provided 
that the local radii 
of curvature of the defect line are large compared 
to $k^{-1}$ at all points along the loop (i.e. the curvature affects 
the correlations of the order parameter 
only on length scales exceeding $k^{-1}$).
We then reach the general result for the structure factor of a vortex
loop in three spatial dimensions 
\begin{equation}
S_{\rm loop}({\bf k})={2 \pi A^{(2)}\over V}{1\over k^4}
\oint ds \, {\delta(k \cos{\theta(s)}) \over \sin^4{\theta(s)}}\,,
\label{loopgen}
\end{equation}
where $s$ denotes the loop arc-length. For example, for a circular loop 
with radius $R$ that has the normal to the loop plane oriented 
at an angle $\zeta$ 
relative to ${\bf k}$, we use the parameterization 
$\theta = \arccos[{\sin({\zeta}) \sin({s/R})}]$ with $s \in [0,2 \pi R]$ 
and obtain 
\begin{equation}
S_{\rm circ}({\bf k})={4 \pi R A^{(2)} \over V}{1\over k^5 \,|\sin{\zeta}|}\,.
\label{loopcirc}
\end{equation}
The $k^{-5}$ dependence in Eq.~(\ref{loopcirc}) agrees with the
general Porod-law form, Eq.~(\ref{genporod}), with the exponent $\chi
= 2d - D = 5$ corresponding to the spatial dimensionality $d=3$ and
defect dimensionality $D=1$.  The result Eq.~(\ref{loopcirc}) is valid
in the range $R^{-1} \ll k \ll \xi^{-1}$, where $\xi$ is the core size
of the disclination.  Note that $S_{\rm circ\/}$ reaches its minimal value,
$4 \pi R A^{(2)}/V k^5$, when ${\bf k}$ lies in the plane of the defect
loop (i.e., $\zeta = \pi/2$), and diverges as the angle $\zeta$
approaches $0$ or $\pi$ (i.e. ${\bf k}$ perpendicular to the loop plane). This
reflects the fact that the length of the loop segments at which the angle
$\theta$ between ${\bf k}$ and the local vortex orientation is
sufficiently close to $\pi/2$ for the segment contribution
Eq.~(\ref{string}) to be nonzero increases with decreasing $\zeta$ (for
$\zeta=0$, we have $\theta=\pi/2$ all along the
loop). Equation~(\ref{loopcirc}) ceases to be valid in the limit $\zeta
\rightarrow 0$; for finite $R$, 
the regime of validity is restricted to angles $|\zeta| > \pi / (k R)$
 \cite{angle_note}.

It is worth stressing that there is no need for any 
large-distance cut-off in the
loop calculations, as none of the integrals leading to
Eq.~(\ref{loopcirc}) diverges at small $k$. The factor of system volume $V$ in
Eqs.~(\ref{string}-\ref{loopcirc}) arises 
purely due to the chosen normalization of the structure factor. 
The presence for each loop segment of an antipodal loop 
segment at separation $2 R$ is reflected in the structure factor
result solely through the fact that the validity of 
Eq.~(\ref{loopcirc}) is restricted to $k$ values larger than $R^{-1}$.

Now consider the situation, typical during 
the phase-ordering process following a quench, 
in which the system contains many vortex loops of random orientations. 
In this situation, the properties of the system are globally isotropic 
and the structure factor 
$S({\bf k})$ depends only on the magnitude $k$ of the scattering wave-vector. 
We may obtain $S(k)$ by averaging the result in Eq.~(\ref{string}) 
over all orientations of the defect segment:
$S_{\rm isotr}(k) \equiv {1/2} \int_0^{\pi} d\theta \, \sin{\theta} \,
S_{\rm seg}(k, \cos{\theta})$.
By using the result (\ref{string}) we obtain the structure factor 
per unit defect length in a globally 
isotropic system:
\begin{equation}
S_{\rm isotr}(k)={\pi A^{(2)} \over V}{1\over k^5}\,.
\label{stringres}
\end{equation}

So far, all arguments in this section have depended 
only on the scattering geometry 
(i.e. the shape and orientation of the defect line and the orientation of the 
wave-vector), and were valid independent 
of the form of the order parameter. 
To obtain specific results for case of the $O(2)$ vector model 
in $d=3$, it suffices to substitute the result 
 $A^{(2)}=4\pi^2$ for the Porod amplitude of the point defect 
in the $O(2)$ vector model in $d=2$ [Eq.~(\ref{EQ:MagRes}) for $d=N=2$] 
into the general expressions (\ref{string}-\ref{stringres}).
For example, using $A^{(2)}=4\pi^2$ in Eq.~(\ref{stringres}), we obtain the 
Porod amplitude in a phase-ordering $O(2)$ vector system in $d=3$ as 
$A^{(3)}=4\pi^3$. This last result is in agreement with the corresponding 
result obtained in Ref.~\cite{Bray-amplitudes}.

\section{Uniaxial nematic systems}
\label{SEC:UniNemSys}

\subsection{The nematic order parameter}
\label{SEC:intronem}

For the case of nematic liquid crystals the local order-parameter 
field $Q_{\alpha\beta}({\bf r},t)$ is a symmetric traceless 
rank-2 tensor with
Cartesian indices $\alpha,\beta=1,2,3$ (see, e.g., Ref.~\cite{de Gennes}).  
In the common case of the {\it uniaxial\/} nematic, the order parameter 
has the form 
\begin{equation}
Q_{\alpha\beta}=
  {3 \over 2} S_1 (u_\alpha u_\beta - {1\over 3} \delta_{\alpha\beta}) \,,
\label{paramQuni}
\end{equation} 
where the order-parameter magnitude $S_1$ determines 
the strength of orientational 
ordering of the nematogen molecules, 
and the director ${\bf u}$ gives the local value of the 
preferred orientation of the molecules. 
In the more general {\it biaxial\/} nematic case, the order parameter may be written as
\begin{equation}
Q_{\alpha\beta}=
  {3 \over 2} S_1 (u_\alpha u_\beta - {1\over 3} \delta_{\alpha\beta}) 
+ {1 \over 2} S_2 (b_\alpha b_\beta - v_\alpha v_\beta) \,,
\label{paramQ}
\end{equation}where 
$\pm {\bf u}$ is the uniaxial director, 
$\pm {\bf b}$ is the biaxial  director, 
${\bf v}\equiv{\bf u \times b}$, and 
the amplitudes $S_1$ and $S_2$ determine, respectively, the strength of 
uniaxial and biaxial ordering. 

 For the nematic, the real-space correlation function 
$C({\bf r},t)$ is defined as
\begin{equation}
C({\bf r}) = 
{1 \over M_{\rm nem}}
\int d^{3}x\,{\rm Tr}\, 
[Q({\bf x})\,Q({\bf x+r})]\,,
\label{EQ:defcrnem}
\end{equation}
where ${\rm Tr}[A B]\equiv A_{\alpha\beta} B_{\beta\alpha}$ 
[cf.~Eq.~(\ref{EQ:defcr})], and the normalization factor $M_{\rm nem}$, 
which enforces $C(r=0)=1$,
is given by
\begin{equation}
M_{\rm nem}=\int d^3x\, 
{\rm Tr}\,[Q({\bf x})\,Q({\bf x})] = {V \over 2} (3 S_1^2 + S_2^2)\,,
\label{normnem}
\end{equation}
in which $V$ is the volume of the system.
For the nematic, Eq.~(\ref{EQ:wigner}) becomes 
\begin{equation}
S({\bf k}) = 
{1 \over M_{\rm nem}} 
{\rm Tr}\,[Q({\bf k})\,Q(-{\bf k})] \,.
\label{wignernem}
\end{equation}
In this Paper, we are restricting our attention to unpolarized 
light scattering, in which case the 
structure factor (\ref{wignernem}) is directly proportional 
to the measured scattered-light intensity \cite{de Gennes}. 

 When present, a given type
of defect will adopt an equilibrium
 configuration of the nematic director that minimizes
the nematic free energy. For future reference, we recall 
that the Frank free energy 
of a {\it uniaxial\/} nematic  has (apart from surface terms) the general
form \cite{de Gennes}
\begin{equation}
E= {1 \over 2} \int d^3x \, \left[ K_{11} (\nabla \cdot {\bf u})^2 +
K_{22} ({\bf u} \cdot \nabla \times {\bf u})^2 +
K_{33} \left| {\bf u} \times (\nabla \times {\bf u}) \right|^2 \right]\,,
\label{genFrank}
\end{equation}
where the splay ($K_{11}$), twist ($K_{22}$), and bend ($K_{33}$) elastic 
constants depend on the order parameter magnitude $S_1$.
Eq.~(\ref{genFrank}) is often simplified 
by adopting the so-called one-constant approximation, 
$K_{11} = K_{22} = K_{33} \equiv K$, in which case the free energy 
(apart from surface terms) can be written as
\begin{equation}
E = {K \over 2} \int d^3x \, (\nabla_i { u}_j) (\nabla_i { u}_j)\,.
\label{oneFrank}
\end{equation}


\subsection{Hedgehog defects in uniaxial nematic systems}
\label{SEC:hedge}

A three-dimensional uniaxial nematic system admits
topologically stable point defects (i.e., nematic hedgehogs) as well as
line defects (i.e., nematic disclinations).  (For a pedagogical 
discussion of the issue of 
topological stability, see, e.g., Ref.~\cite{Mermin}.) 
The one-constant approximation to the Frank free energy, 
Eq.~(\ref{oneFrank}), admits 
(up to global rotations) two minimum-energy point-defect 
solutions with unit topological 
charge.
In the {\it radial\/} hedgehog configuration, the director 
$\pm {\bf u}$ points everywhere radially outwards from the center of
the defect, and is given by ${\bf u} = {\bf r}/r$. The {\it hyperbolic\/}
hedgehog configuration can be obtained from the radial hedgehog configuration
by inverting one of the components of the director ${\bf u}$, e.g. 
${\bf u} = (-x/r,y/r,z/r)$. Note that the two configurations are 
topologically
equivalent \cite{Mermin}.

 The calculation described
below gives identical results for both the radial and the hyperbolic hedgehog;
for simplicity, we shall work with the radial hedgehog configuration of 
the director,
\begin{equation}
Q_{\alpha\beta}({\bf r}) = 
{3 \over 2} S_1\, 
\left({r_\alpha \over r}{r_\beta \over r} 
	-{1 \over 3}\delta_{\alpha\beta}\right)\,.
\label{hedge}
\end{equation}

The following calculation of $S({\bf k})$ closely mirrors the steps
Eqs.~(\ref{pointsk})--(\ref{EQ:MagRes}) in our $O(N)$ vector model
calculation.  First, we observe that the Fourier transform of
$Q_{\alpha\beta}({\bf r})$ is itself a symmetric, traceless, rank-2 tensor, and
that due to symmetry it can depend only on the direction $\pm {\bf k}$.
The general form of such a tensor is
\begin{equation}
Q_{\alpha\beta}({\bf k})=
{3\over{2}}S_{1}
\left(k_\alpha k_\beta-{1 \over 3}\delta_{\alpha\beta}\, 
k^2)\,g(k^{2}\right) \,,
\label{hedgesk}
\end{equation}
where $g(k^{2})$ is a  scalar function of the scalar $k^{2}$.
By contracting Eq.~(\ref{hedgesk}) with $k_\alpha k_\beta$ and 
solving for $g$ we can rewrite Eq.~(\ref{hedgesk}) as 
\begin{eqnarray}
Q_{\alpha\beta}({\bf k})
&=&
{9S_{1}\over{4k^{4}}}
 \left(
k_\alpha k_\beta-{1\over 3}\delta_{\alpha\beta}\,k^2
\right)
\int d^{3}r\,
\left[
\left(
{{\bf k}\cdot{\bf r}\over{r}}
\right)^{2}
-{k^{2}\over{3}}\right]
{\rm e}^{i{\bf k}\cdot{\bf r}}
\nonumber
\\
&=&
-{9S_{1}\over{4k^{4}}}
 \left(
k_\alpha k_\beta-{1\over 3}\delta_{\alpha\beta}\,k^2
\right)
\left[
 (2\pi)^{3}{k^{2}\over{3}}\delta({\bf k})
+{\partial^{2}\over{\partial\lambda^{2}}}\bigg\vert_{\lambda=1}
G(\lambda^{2}k^{2})
\right]\,.
\label{EQ:WithG}
\end{eqnarray}
In this expression the 
$\delta$-function part corresponds to the forward-scattered
beam, and will be dropped henceforth, as it does not influence the
large-$k$ behavior.  The function $G$ is defined via
\begin{equation}
G(\lambda^{2}k^{2})\equiv
\int d^{3}r\,r^{-2}\,
{\rm e}^{i\lambda{\bf k}\cdot{\bf r}}\,, 
\end{equation}
i.e., $G$ is the three-dimensional Fourier transform of the 
potential $r^{-2}$, which can readily be shown to be given by 
\begin{equation}
G(\lambda^{2}k^{2})=
{2\pi^{2}\over{\lambda k}}\,.
\end{equation}
Evaluating the second derivative of $G$, and inserting it into 
Eq.~(\ref{EQ:WithG}), leads to 
\begin{equation}
Q_{\alpha\beta}({\bf k})
=-{9\pi^{2}S_{1}\over{4k^{5}}}
 \left(k_\alpha k_\beta-{1\over 3}\delta_{\alpha\beta}\,k^2\right)\,.
\label{EQ:Qans}
\end{equation}
 From Eq.~(\ref{normnem}) we see that the normalization 
factor $M_{\rm nem}$ 
takes the value $3VS_{1}^{2}/2$.  Combining 
Eqs.~(\ref{normnem}), (\ref{wignernem})
 and (\ref{EQ:Qans}) gives for the structure factor
$S(k)$ of a single nematic hedgehog defect in a system of volume $V$
 the result 
\begin{equation}
S(k)={36\pi^{4} \over V}{1\over{k^{6}}}\,.
\label{finalhedge}
\end{equation}
Note that the Porod amplitude $36\pi^{4}$ for the nematic hedgehog 
differs substantially from the Porod amplitude $12 \pi^{3}$ 
(see Sec.~\ref{SEC:PDinOnSymm})
for the $O(3)$ monopole in $d=3$.
\subsection{Disclination lines in uniaxial nematic systems}
\label{SEC:DisInUni}

We now calculate $S({\bf k})$ for the 
case of a disclination line defect in a uniaxial nematic system. 
In the topologically stable disclination configuration, 
the director rotates about the core of the disclination  
by $\pm 180^\circ$ (see, e.g., Ref.~\cite{Mermin}). 
Disclinations with $\pm 360^\circ$ rotations are unstable 
towards the ``escape in the third dimension'' 
\cite{Mermin} and do not have a singular core. 
We shall concentrate our attention on the topologically 
stable disclinations (appearing as thin threads in optical 
observations) as the  
$\pm 360^\circ$ disclinations (appearing as thick threads) 
do not give raise to a power-law 
contribution to $S(k)$.

In Section \ref{SEC:VortexLines}, we gave general geometric 
arguments that allowed us to express the structure factor of a linear defect 
in $d=3$ in several specific configurations 
in terms of the Porod amplitude $A^{(2)}$ of the point defect 
in $d=2$ (obtained as the cross-section of the line defect in $d=3$). 
Equations~(\ref{string}--\ref{stringres}) are valid regardless of the
order parameter in question and may, thus, also be used for
nematic systems. The remaining task therefore is to obtain an expression 
for $A^{(2)}$ in the nematic case. 

As the director
rotates about the core by $180^\circ$, it is clear that 
the approach of Sec.~\ref{SEC:hedge}, which is based on rotational
symmetry, cannot be directly applied.  It is readily seen,
however, that the calculation of the Porod-law amplitude for the
$180^\circ$ nematic defect can be mapped on to the corresponding
calculation for the $360^\circ$ defect in the $O(2)$ vector model in $d=2$.
For the $180^\circ$ uniaxial nematic defect, in the one-constant 
Frank free energy approximation, the order-parameter configuration
is given (up to a global rotation) by Eq.~(\ref{paramQuni}) with  
${\bf u}({\bf x})=\big(\cos{1\over{2}}\phi({\bf
x}),\sin{1\over{2}}\phi({\bf x}),0\big)$,
where $\phi({\bf x})$ is the polar angle of the radius-vector 
${\bf x}$ in the plane perpendicular to the disclination line.
The correlation function Eq.~(\ref{EQ:defcrnem}) then becomes 
\begin{mathletters}
\begin{eqnarray}
C^{(2)}_{180^\circ}(r)
&=&
{(3 S_1/2)^2\over M_{\rm nem}}\int d^2x\, 
\cos^2\left({1\over{2}}\phi({\bf x})-{1\over{2}}\phi({\bf x+r})\right)+ c_1  
\\
&=&
{(3 S_1/2)^2\over M_{\rm nem}}\int d^2x\,{1\over 2} 
\cos\left(\phi({\bf x})-\phi({\bf x}+{\bf r})\right) + c_2 \,,
\label{EQ:map}
\end{eqnarray}
\end{mathletters}
where 
$c_1$ and $c_2$ are numerical constants. 
On the other hand, for the $360^\circ$ defect in the $O(2)$ model in
$d=2$, the order-parameter configuration is given by
${\bf \Phi}({\bf x}) = \big(\cos\phi({\bf x}),\sin\phi({\bf x})\big)$,
and the corresponding correlation
function, Eq.~(\ref{EQ:defcr}), is given by
\begin{equation}
C^{(2)}_{360^\circ}(r)={1 \over M_{O(2)}} 
\int d^2x\,\cos[\phi({\bf x})-\phi({\bf
x}+{\bf r})]\,.
\label{EQ:map2}
\end{equation}
After Fourier-transforming Eqs.~(\ref{EQ:map}) and (\ref{EQ:map2}) and
omitting the $\delta$-function term arising from the constant $k_2$ in
Eq.~(\ref{EQ:map2}), we obtain that the structure factors of the two
defects are related simply by 
\begin{equation}
S^{(2)}_{180^\circ}(k)
={(3S_1/2)^2 M_{O(2)}\over 2M_{\rm nem}} S^{(2)}_{360^\circ}(k)\,.
\label{EQ:map3}
\end{equation}
Using the result $S^{(2)}_{360^\circ}(k)= 4 \pi^2 k^{-4} / {V^{(2)}}$ [i.e.,
Eq.~(\ref{EQ:MagRes}] for $d=2$, with the area of the system
denoted by $V^{(2)}$), and the normalization factors
$M_{O(2)}=V^{(2)}$ and $M_{\rm nem}=V^{(2)} 3S_1^2/2$, we finally obtain 
the structure factor $S^{(2)}_{180^\circ}(k)$ for 
the $180^\circ$ point defect in a two-dimensional nematic:
\begin{equation}
S^{(2)}_{180^\circ}(k) = {3\pi^2 \over {V^{(2)}}}  {1 \over k^4}\,.
\label{EQ:2d180}
\end{equation}
By using this result (i.e., $A^{(2)} = 3\pi^2$) in Eq.~(\ref{loopcirc}), 
we obtain 
the structure factor of a nematic disclination loop of radius~$R$
\begin{equation}
S_{\rm circ\/}(k)={12 \pi^3 R \over V} {1\over k^5 \,|\sin{\zeta}|}\,.
\label{EQ:disclcirc}
\end{equation}
Likewise, by using Eq.~(\ref{stringres}) we obtain the orientation-averaged contribution 
to the structure factor from a segment of length $L$
of a nematic disclination
\begin{equation}
S(k)={L \over V}\, {3\pi^3} {1\over k^5}\,.
\label{EQ:finaldiscl}
\end{equation}
It follows that
the structure factor 
for a globally isotropic system containing disclination lines 
is given by $S(k) = \rho\, {3\pi^3} k^{-5}$, with
the defect density $\rho$ having the meaning of the {\it total
disclination length per unit volume of the system\/}. 

\begin{figure}
\centerline{\epsfig{figure=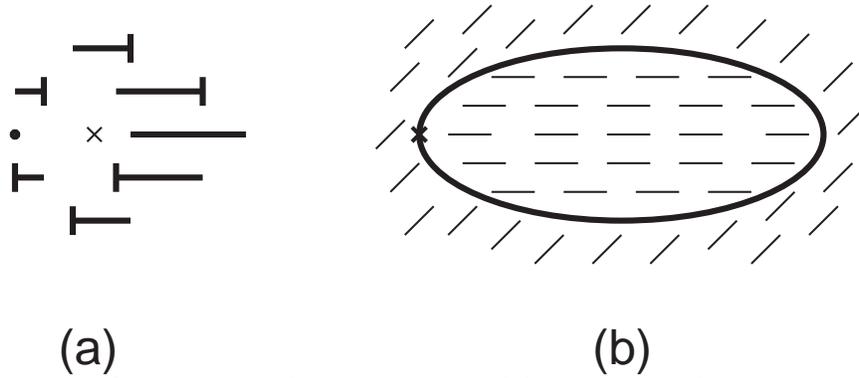,angle=0,width=4.5in,silent=}}
\caption{Order parameter configuration around a twist-type ring defect
in a uniaxial nematic system. (a) Cross-section of the director
configuration around a disclination segment running perpendicular to
the page with the core passing through the point ``x''. 
The ``nail head'' marks the end of the director that lies in
front of the page. The director rotates by $180^{\circ}$ around the
core of the disclination; the rotation axis is vertical, i.e., {\it
perpendicular\/} to the orientation of the disclination segment.  (b)
The director configuration in the plane of the twist disclination
loop. The director outside the disclination loop is rotated by
$90^{\circ}$ with respect to the director inside the loop; the
rotation axis is everywhere perpendicular to the loop plane. The cross-section
shown in (a) corresponds to the configuration around the disclination
segment marked by ``x'' in (b).
\label{twistloop}}
\end{figure}

\begin{figure}
\centerline{\epsfig{figure=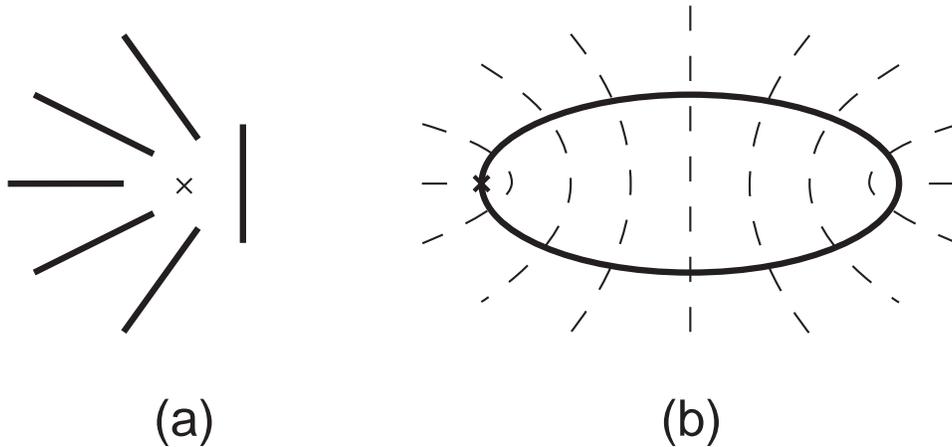,
width=5in,angle=0,silent=}}
\caption{Order-parameter configuration around a wedge-type ring defect
in a uniaxial nematic system. (a) Cross-section of the director
configuration around a disclination segment running perpendicular to
the page with the core passing through the point ``x''. 
The director rotates by $180^{\circ}$ around the core of the
disclination, with the rotation axis {\it parallel\/} to the
orientation of the disclination segment.  (b) The director
configuration of the wedge disclination loop.  The director is
perpendicular to the loop plane everywhere inside the loop.  Outside
of the loop (in any plane), the director adopts the radially
symmetrical configuration of a nematic hedgehog.  The cross-section
shown in (a) corresponds to the configuration around the disclination
segment marked by ``x'' in (b).
\label{wedgeloop}}
\end{figure}


\subsection{Ring defects in uniaxial nematic systems}
\label{SEC:RingsInUni}

Topologically stable disclinations lines in a uniaxial nematic cannot
terminate in the bulk --- they must form closed loops, 
or else extend to the system boundary. Depending on the details of the director
configuration, the disclination loop can carry any (integral)
monopole charge \cite{Nakanishi}. In this section, we discuss typical 
examples of 
disclination loops with zero and nonzero monopole charges 
and their contributions to the structure factor.

Consider first the case of a ``twist disclination'' loop, which carries zero 
monopole charge. The
cross-section of the director configuration around a twist
disclination is shown in Fig.~\ref{twistloop}a. The axis of rotation
of the director is {\it perpendicular\/} to the direction of the
disclination line. The director configuration obtained by forming a
closed loop of a disclination that locally has the ``twist''
structure of Fig.~\ref{twistloop}a is illustrated in Fig.~\ref{twistloop}b. 
This director configuration is homogeneous at large distances
from the loop, and consequently does not result in a power-law
contribution to the structure factor for $k \ll R^{-1}$.  
For length scales smaller than $R$ (i.e., for $k \gg R^{-1}$), the twist 
disclination loop is characterized by the structure factor given 
in Eq.~(\ref{EQ:disclcirc}). 

\begin{figure}[hbt]
\centerline{\epsfig{figure=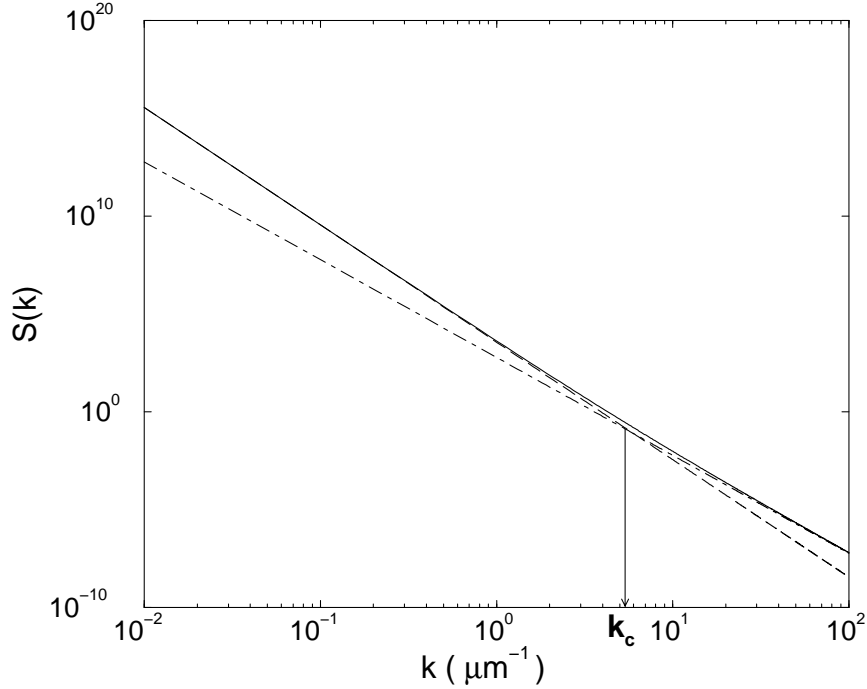,
width=4.5in,angle=270,silent=}}
\caption{Structure factor $S({\bf k})$ of the ring defect
configuration in Fig.~\protect\ref{wedgeloop}. The scattering 
wave-vector ${\bf k}$ is assumed to be oriented at an angle $\zeta \neq
0$ with the respect to the ring axis.  The two Porod
regimes, arising for length scales larger or smaller than
the ring radius $R$, are shown as the dot-dashed line [$S(k)= 
36 \pi^4 / (V k^{6})$] and the dotted line [$S(k)= 12 \pi^3 R /(V k^{5}
|\sin{\zeta}|)$], respectively. The solid curve schematically shows
the crossover between the two power-law regimes.  The crossover
wave-vector $k_{\rm c}$, as defined in the picture, is given by $k_c=3 \pi 
R^{-1} |\sin{\zeta}|$. Thermal fluctuations can further modify the form of
$S(k)$ shown in the figure, if the distance to the nearest other defect
or the system boundary is large; these effects are discussed in
Sec.~\protect\ref{SEC:thermal}.
\label{fig2}}
\end{figure}

Next, consider the ``wedge disclination'' loop. In this 
configuration, the director rotates about an axis {\it parallel\/} to
the disclination line (see Fig.~\ref{wedgeloop}a). The resulting director
configuration (see Fig.~\ref{wedgeloop}b) has the structure of a radial
hedgehog (see Sec.~\ref{SEC:hedge}) outside of the disclination loop.
This results in a power-law contribution to the structure factor 
characterizing the nematic hedgehog
in $d=3$, Eq.~(\ref{finalhedge}), valid for $k<R^{-1}$, where $R$ 
is the ring radius. On the other hand, for $k \gg R^{-1}$, 
we have the usual disclination-loop structure factor, 
Eq.~(\ref{EQ:disclcirc}). To summarize, 
the structure factor of a wedge disclination loop is given by 
\begin{eqnarray}
S(k)=\cases{ {36 \pi^4 \over V}{1 \over k^6},& $k \leq R^{-1}$;\cr
             {12 \pi^3 R \over V} {1\over k^5 \,|\sin{\zeta}|},& $R^{-1} \ll 
                                              k \ll \xi^{-1}$\,.}
\label{wedgering}
\end{eqnarray}
Therefore the large-$k$ region of the structure factor measured in a
system containing a wedge disclination loop of radius size $R$ 
is expected to exhibit a crossover
between two power-law regimes with exponents $6$ and $5$ 
(Fig.~\ref{fig2}). The crossover point, as defined in Fig.~\ref{fig2} ,
is predicted to occur at the value of $k_{\rm c}=3 \pi \zeta /R$.  
Note that measuring the location $k_{\rm c}$ of the crossover 
allows one to obtain information about the ring-defect size $R$ from {\it
un-normalized\/} data for $S(k)$; this should be contrasted to the 
use of the Porod law based on 
Eq.~(\ref{EQ:finaldiscl}), 
where the total string length is extracted from
the (rarely experimentally available) normalized structure factor. 

Even though the wedge disclination loop is a topologically 
admissible configuration, the (energetics-dependent) question 
of its occurrence in real nematic systems  
has recently attracted some interest. 
The energetic stability of the wedge loop configuration was
investigated theoretically in Refs.~\cite{Mori} and \cite{Lavrentovich}.  
There, it
was found that provided that certain restrictions on the elastic
constants in the nematic free energy are satisfied, there does indeed exist a
non-zero equilibrium radius of the loop: a
ring of a larger (smaller) radius will tend to shrink (expand).  Depending
on the values of the elastic constants,
Ref.~\cite{Lavrentovich} predicts $R_{\rm eq}$ in the range
$10\xi-10^4\xi$, where $\xi$ is the coherence length of the nematic order 
parameter 
(i.e., $\xi$ is of the order of $10-100\,\AA$). Note that in a nematic with
a positive dielectric or diamagnetic susceptibility anisotropy,
the loop can be made unstable towards expansion to a larger radius
by applying an electric or magnetic field 
along the axis of the defect ring \cite{Bondar91}.
Wedge disclination loops with a non-zero 
equilibrium radius were recently observed in numerical simulations 
\cite{Toyoki94}. Experimental observations of such a configuration, 
however, are currently not available. The form of the structure factor 
predicted in Eq.~(\ref{wedgering}), with its characteristic crossover, 
may be used as an experimental signature for the observation 
of the wedge ring configuration \cite{sattelite}. 


\subsection{Thermal fluctuations of director orientation}
\label{SEC:thermal}

Up to this point, we have ignored thermal fluctuations --- we have effectively 
assumed that the structure factor was measured at zero temperature.
 Nematic liquid crystalline phases, however,
typically occur and are studied in the room-temperature range, 
and thermal effects
can play an important role the scattering of light. 
Due to the presence of low-energy fluctuations
in the orientation of the nematic director, 
the system exhibits enhanced turbidity 
throughout the whole range of temperatures where the nematic phase is stable.
These transverse fluctuations of the director 
result in a power-law contribution
to the structure factor $S(k)$, even when no defects are present in the system.

We now very briefly review the theory of scattering by thermal fluctuations in
bulk uniaxial nematic liquid crystals \cite{de Gennes,twodimthermal}. 
Consider a well-aligned
region of the sample with the average director ${\bf u_0}$ parallel 
to the $z$-axis. The local value of the director can be written
(to first order in ${\bf u_{\perp}}$) as ${\bf u}({\bf r})
 = {\bf u_0} + {\bf u_{\perp}}({\bf r})$,
where the transverse fluctuation ${\bf u_{\perp}}({\bf r})$, 
by definition, satisfies ${\bf u_{\perp}}({\bf r})\perp{\bf u_0}$ and 
we assume that $|{\bf u_{\perp}}({\bf r})| \ll 1$. 
The nematic structure factor, Eq.~(\ref{wignernem}), then becomes 
[upon dropping terms proportional to $\delta(k)$]
\begin{equation}
S(k) = {3 \over 2 V}  
{\bf u_{\perp}}({\bf k}) \cdot {\bf u_{\perp}}({\bf -k})\,.
\label{skfluct}
\end{equation}
Note that the quantity ${\bf u_{\perp}}({\bf k}) \cdot {\bf u_{\perp}}({\bf -k})$ 
must now be regarded as a thermal average. It's value is readily obtained from 
the equipartition theorem. In the ``one-constant'' approximation,
using ${\bf u}({\bf r})
 = {\bf u_0} + {\bf u_{\perp}}({\bf r})$
in Eq.~(\ref{oneFrank}) results in the free energy
\begin{equation}
E = {1 \over 2} K \int {d^3k \over (2 \pi)^3} 
k^2 |{\bf u_{\perp}}({\bf k})|^2\,.
\label{harmonic}
\end{equation}
Furthermore, 
it is possible to decompose ${\bf u_{\perp}}({\bf k})$ into two 
orthogonal modes
that decouple \cite{de Gennes}; by equipartition, we then have 
$|{\bf u_{\perp}}({\bf k})|^2 = k_{\rm B} T V$. 
Equation~(\ref{skfluct}) thus becomes
\begin{equation}
S_{\rm T}(k) = {3 \over 2}  {k_{\rm B} T \over K k^2}\,.
\label{EQ:thermal}
\end{equation}

Taking a typical value $K = 10^{-6}$ dynes 
for the elastic constant and
$k_{\rm B} T \simeq 0.5\cdot 10^{-13}$ erg for the thermal energy at room 
temperatures, one obtains the estimate 
\begin{equation}
S_{\rm T}(k) \simeq {8 {\AA} \over  k^2} \,.
\label{EQ:thermalnumer}
\end{equation}
Note that in the light-scattering literature, 
the structure factor is often defined without the normalization
factor $M_{\rm nem}$ in Eq.~(\ref{wignernem}), 
and then has the dimension of volume squared,
rather than of volume as in our case. 

It should be stressed that Eq.~(\ref{EQ:thermal}) was derived 
from a free energy appropriate for small fluctuations in a 
well-aligned part of the nematic
sample. Equation~(\ref{EQ:thermal}) can therefore be used 
to describe scattering 
from a system with a spatially varying director only provided that  
there are no appreciable variations in the
static director orientation on the scale of $k^{-1}$.

Even in strongly inhomogeneous samples containing topological defects, 
thermal fluctuations will dominate the scattering intensity  
provided either that 
the defect density is sufficiently low or that $k$ is sufficiently high. 
To illustrate this point, 
we now consider light scattering from a nematic droplet of radius
$R$ with normal (homeotropic) boundary conditions on the droplet surface. 
In the minimum-energy
configuration, the droplet contains a radial hedgehog in the
center. (As discussed 
in Sec.~\ref{SEC:RingsInUni},
the radial hedgehog is, in fact, expected to take the form of a wedge
disclination loop of small radius $r$; in the present paragraph, we restrict 
our attention to length scales exceeding $r$ and do not consider the detailed 
structure of the hedgehog.) 
Consequently, for a given scattering wave-vector $k \gg 2 \pi /
R$, the central region of the droplet will contribute to the structure
factor according to Eq.~(\ref{finalhedge}), $S_{\rm def}(k) 
= V^{-1} 36 \pi^4  k^{-6}
$, where $V=4 \pi R^3/3$.
 In the region $2 \pi / k < r < R$, the static order-parameter
configuration does not contain variations that 
contribute substantially to $S(k)$; 
however, thermal fluctuations on the scale $2
\pi / k$ still exist. The contribution to $S(k)$ from these is
given by Eq.~(\ref{EQ:thermalnumer}). The resulting ratio
$\gamma(k)$ of thermal and static scattering intensities, i.e. $\gamma(k) 
\equiv S_{\rm T}(k) / S_{\rm def}(k) = 0.04 (R k)^3 k$, where $k$ is given
in inverse \AA, can very in a wide range. Clearly the static
contribution $S_{\rm def}(k)$ dominates when $k$ is sufficiently small
(i.e. near the forward-scattered beam). 
Consider, on the other hand, scattering at the 
experimentally accessible value of $k=10^{-5} \AA^{-1}$. 
For a droplet
of radius $R=1\, {\rm cm\/}$, we obtain $\gamma \simeq 400$; 
for a droplet
of smaller radius $R=100 \,\mu {\rm m}$, we obtain $\gamma \simeq 0.0004 $.
We see that either the thermal or the static contribution to the scattering
intensity can completely dominate 
for $R$ and $k$ in the experimentally reasonable range.
(For the sake of completeness we note that we have not taken 
into account scattering 
from the droplet surface in the arguments given above.) 

\subsection{Porod tails in the phase-ordering kinetics of uniaxial nematics}
\label{SEC:exper}

We are now ready to discuss in detail the short-distance 
form of the structure factor arising 
during the phase-ordering process in uniaxial nematic 
liquid crystals, and to relate
our conclusions to the experimental results reported in 
Refs.~\cite{Wong92,Wong93}.

We first briefly recall some general features of phase-ordering
kinetics following a quench from a disordered to an ordered phase
\cite{Gunton83,Bray-review}.  Phase-ordering kinetics initially
attracted theoretical attention due to the property of dynamical
scaling of the order-parameter correlations at late times after the
quench. In its simplest form, the corresponding 
dynamical scaling hypothesis states
that all time-dependent length scales in the system have the same
asymptotic time dependence (and, consequently, one can define a single 
``characteristic'' length scale). 
In most systems with non-conserved order-parameter 
dynamics, this time-dependence is given by the power law
$L(t) \propto t^{1/2}$. In systems
containing topological defects, the characteristic length scale $L(t)$
can usually be extracted as the typical defect-defect 
separation at time $t$.
 
The process of phase ordering has been successfully studied
experimentally in systems described by a scalar order parameter (e.g.,
in binary alloys---see Ref.~\cite{alloys}).  However, analogous
experiments were found to be difficult to perform in most systems with
continuous symmetries (such as ferromagnets or liquid He$^4$).  
In the recent years,
a series of experiments \cite{Chuang91,Chuang93,Wong92,Wong93} have
demonstrated that nematic liquid crystals provide a system in which
phase ordering (following a quench from the isotropic to the nematic
phase) is readily accessible to experimental investigation. 
In Refs.~\cite{Wong92,Wong93}, the phase-ordering process was 
studied by measuring the time-dependent nematic structure factor 
$S(k,t)$. The present section includes a discussion of 
 the large-$k$ region of 
the structure factor measured in these experiments; however, we also discuss  
effects that may be observed under different experimental 
conditions. 


As we saw in Secs.~(\ref{SEC:hedge})--(\ref{SEC:RingsInUni}), 
a uniaxial nematic liquid crystalline system can contain,
simultaneously, point defects (hedgehogs) and line defects
(disclinations) \cite{textures}.  If the average separation between 
hedgehogs is given by $L_{\rm hedg}$ and that between 
disclinations by $L_{\rm discl}$, a unit volume of the system contains
$\rho_{\rm hedg} = L_{\rm hedg}^{-3}$ hedgehogs and a length 
$\rho_{\rm discl} =
L_{\rm discl}^{-2}$ of disclinations.  Now consider scattering at a
wave-vector $k$ such that $k^{-1} \ll L_{\rm discl}$ and $k^{-1} \ll
L_{\rm hedg}$.  The total structure factor per unit volume due to 
hedgehogs is then given by $\rho_{\rm hedg} \, 36\pi^{4} /k^6$ 
[see Eq.~\ref{finalhedge}], and that due to disclinations by 
$\rho_{\rm discl} \, {3\pi^3} /k^5$ [see
Eq.~(\ref{EQ:finaldiscl})]. (Here we have assumed that the disclination 
lines are predominantly of the twist type, as is expected from 
energetic considerations.) In the parts of the system
that are not within distance $k^{-1}$ from the nearest
disclination or hedgehog (which, by assumption, is most of the volume
of the system), the static director configuration is effectively
homogeneous on scales of order of $k^{-1}$.  Consequently, 
Eq.~(\ref{EQ:thermal})
for the strength of scattering due to thermal fluctuations applies,
and per unit volume, we have $S_{\rm T}(k)  \simeq 8 {\AA} / k^2$.  (Notice
that the length scale occurring in the expression for $S_T(k)$ is
microscopic, as opposed to the defect separations occurring in the
hedgehog and disclination contributions.)  The total structure factor is
therefore given by
\begin{equation}
S(k) = \rho_{\rm hedg} \,  {36\pi^{4} \over k^6} + \rho_{\rm discl} \, 
 {{3\pi^3} \over k^5} +
 {8 {\AA} \over k^2} \,.
\label{combine}
\end{equation} 

First consider a system in which disclinations dominate over hedgehogs.
We now estimate, for a given density of disclinations $\rho_{\rm discl}$ 
(and $\rho_{\rm hedg}=0$),
the range of scattering wave-vectors $k$
in which the the generalized Porod law behavior, $S(k) \propto k^{-5}$, 
can be observed. The lower limit on the range of $k$ is given by the inverse average
separation of defects: $k_{\rm l}=L_{\rm discl}^{-1}=\rho_{\rm discl}^{2}$.
The upper limit $k_{\rm u}$ is given by the $k$ value for which $S_{\rm T}(k) = 8\, {\AA} k^{-2}$
exceeds the Porod law contribution, $S_{discl}=\rho_{\rm discl} 3 \pi^3 k^{-5}$. 
This gives $k_{\rm u} = (3 \pi^3 / 8 {\AA})^{1/3} L_d^{-2/3}$. 
We see that although both $k_{\rm l}$ and $k_{\rm u}$ decrease when 
the separation of defects $L_{\rm discl}$ increases, the ratio $k_{\rm u}/k_{\rm l}$ {\it increases\/} 
as $(L_{\rm discl} / 0.09 {\AA})^{1/3}$. Analogous estimates for observing
the $S(k) \propto k^{-6}$ behavior in a system dominated by hedgehogs 
yield $k_{\rm l}=L_{\rm hedg}^{-1}=\rho_{\rm hedg}^{3}$, 
$k_{\rm u} = (36 \pi^4 / 8 {\AA})^{1/4} L_{\rm hedg}^{-3/4}$,
and $k_{\rm u}/k_{\rm l} =(L_{\rm hedg} / 0.002 {\AA})^{1/4}$.
 Consequently, the Porod form of the structure factor
 is predicted to extend over one to three decades
in $k$ for the experimentally accessible values of defect
separation of the order of $0.1 \,\mu {\rm m}$ to $1\, {\rm cm}$, 
for either a disclination-dominated 
or a hedgehog-dominated system \cite{field}. 
Crossover effects between $k^{-6}$ and $k^{-5}$ scaling
in situations when both hedgehogs and disclinations are present
may further diminish the range in which these 
power laws can be clearly observed. (A specific example 
of such behavior will be seen below.)

It should be noted that measuring the location of the crossover
wave-vector $k_{\rm u}$ permits the efficient estimation of the density of
defects in the system.  Specifically, if one measures the location of
the crossover $k_{\rm u}$ in a disclination-dominated system, the total
disclination length per unit volume (expressed in units of ${\AA}^{-2}$) is
given by $(2.3 / k_{\rm u})^{3/2}$. In a hedgehog-dominated system, the
number of hedgehogs per unit volume (expressed in units of ${\AA}^{-3}$) is
given by $(4.6 / k_{\rm u})^{4/3}$. Although it is also possible to estimate
the defect density from the position $k_{\rm l}$ of the crossover at low
wave-vectors, it should be noted that this crossover is less steep than
the crossover at $k_{\rm u}$, and its precise character depends on the nature
of the correlations amongst the defects.

We now relate our results to the measurements of $S(k)$ in a uniaxial
nematic undergoing phase ordering in the experiments of Wong et
al.~\cite{Wong92,Wong93}.  The range of scattering wave-vectors
investigated in Refs.~\cite{Wong92,Wong93} was $k=1000
{\rm cm}^{-1}$--$10000 {\rm cm}^{-1}$, and the typical defect separation in the
system at intermediate and late times after the quench was of the
order of $10 \,\mu {\rm m}$ (see Fig.~1 in Ref.~\cite{Wong92}).  Direct
optical observations in this and related \cite{Chuang91,Chuang93}
phase-ordering systems indicate that disclinations dominate over
hedgehogs; the exact relative proportion of their densities is, however,
difficult to measure.

\begin{figure}[hbt]
\centerline{\epsfig{figure=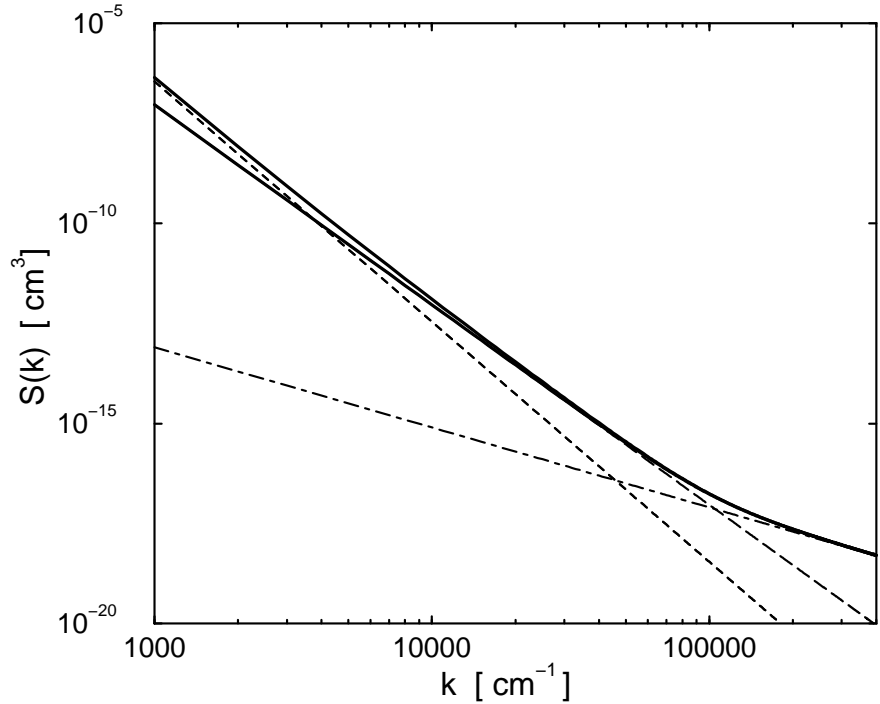,
width=4.5in,angle=270,silent=}}
\caption{
Theoretically predicted short-distance structure factor $S(k)$ for a 
3-dimensional uniaxial 
nematic system undergoing phase ordering. 
The lower solid curve
shows $S(k)$ for disclination density 
$\rho_{\rm discl}=10^6 cm^{-2}$ and hedgehog density zero.
The upper solid curve corresponds to disclination density 
$\rho_{\rm discl}=10^6
cm^{-2}$ and hedgehog density $\rho_{\rm hedg}=10^8 cm^{-3}$.  
The long-dashed, dashed, and dot-dashed lines show the contributions 
due to disclinations, hedgehogs, and thermal fluctuations, respectively.}
\label{FIG:Wong}
\end{figure}

In Fig.~\ref{FIG:Wong}, we plot the structure factor $S(k)$ predicted
by Eq.~(\ref{combine}) for two situations in which the average separation
between defects is of the order of $10 \,\mu {\rm m}$. The lower solid curve
shows $S(k)$ for $\rho_{\rm discl}=10^6 \,{\rm cm}^{-2}$ and $\rho_{\rm hedg}=0$
(i.e., a configuration dominated by disclinations with $L_{\rm discl}=10
\,\mu {\rm m}$). The upper solid curve corresponds to $\rho_{\rm discl}=10^6
\,{\rm cm}^{-2}$ and $\rho_{\rm hedg}=10^8 \,{\rm cm}^{-3}$ 
(i.e., the system 
contains, in addition, 
 hedgehogs with average spacing $L_{\rm hedg} = 22 \,\mu {\rm m}$).  We
restrict the plot to the region $k > (10 \,\mu {\rm m})^{-1} = 10^3 \,{\rm cm}
^{-1}$,
where Eq.~(\ref{combine}) is expected to be applicable.  The broken
lines in Fig.~\ref{FIG:Wong} show the three individual contributions
to $S(k)$ arising from hedgehogs, disclinations, and thermal
fluctuations.

It is seen that for both $\rho_{\rm hedg}=0$ and $\rho_{\rm hedg}=10^8
\,{\rm cm}^{-3}$, the crossover to the thermal-fluctuation-dominated regime
[$S(k) \propto k^{-2}$] occurs at approximately $k = 50000 \,{\rm cm}^{-1}$,
i.e., well beyond the range of $k$ measured in
Refs.~\cite{Wong92,Wong93} and at the limit of the range of $k$
accessible, in principle, with visible light.  Indeed, the structure-factor 
data in Refs.~\cite{Wong92,Wong93} do not show any sign of such
a crossover. It should be noted, however, that if the scattering
experiments would have been 
performed at a (still later) time when the average defect
separation would have reached $300 \,\mu {\rm m}$, 
the crossover value would have decreased to
$k \simeq 5000 \,{\rm cm}^{-1}$, and the sharp crossover to the thermal regime
would have been clearly visible with the experimental setup of
Refs.~\cite{Wong92,Wong93} provided that the scattering probe were 
sufficiently sensitive.

In the case $\rho_{\rm hedg}=0$, the structure factor in
Fig.~\ref{FIG:Wong} exhibits the Porod behavior $S(k) \propto k^{-5}$,
characteristic of disclinations, over a range of 1.5 decades ($k=1000
\,{\rm cm}^{-1}$ to $k=50000 \,{\rm cm}^{-1}$).  
In the case $\rho_{\rm hedg}=10^8\,
{\rm cm}^{-3}$, a crossover between $S(k) \propto k^{-6} $ and $S(k) \propto
k^{-5}$ is spread between $k=1000 \,{\rm cm}^{-1}$ 
and $10000 \,{\rm cm}^{-1}$, and
leaves only a very narrow range of $k$ for which the $S(k) \propto k^{-5}$
scaling behavior may be observed.  The fact that hedgehogs result in a
substantial modification of the structure factor even when $L
\rho_{\rm hedg} / \rho_{\rm discl} = 0.1$ is rather surprising, and is due to
the large ratio, $A_{\rm hedg}/A_{\rm discl} = 12 \pi$, of the Porod
amplitudes for hedgehogs and disclinations.

The structure-factor data reported in Refs.~\cite{Wong92,Wong93} (for
three-dimensional samples) were fit in these references to a power-law
with exponent $6 \pm 0.3$. Some of these data (specifically Fig.~7 in
Ref.~\cite{Wong93}) were later re-analyzed in
Ref.~\cite{Bray-nematic}, with the conclusion that the asymptotic
slope was closer to $5$, but was approached from {\it above \/},
i.e., through effective exponents lying between $5$ and $6$. Our results show
(see the upper solid line in Fig.~\ref{FIG:Wong}) that such behavior
is expected to arise if a sufficient number of hedgehogs is present in
the system.  The relative density of disclinations and hedgehogs was
not examined in the experiments of Refs.~\cite{Wong92,Wong93}.  Other
experiments investigating phase ordering in nematics, however,
indicate that hedgehogs are rare compared to disclinations. 
(For the sake of completeness
we note that it is possible to prepare nematic systems dominated by
hedgehog defects: thus, in Ref.~\cite{Pargelis96}, a liquid crystalline
system was quenched from the isotropic to the isotropic-nematic
biphasic region, and a large number of hedgehogs formed upon
coalescence of the nematic droplets.)
Specifically, it was found in Ref.~\cite{Chuang93} that hedgehogs occurred
in significant numbers only within a specific intermediate range of
times after the quench, and that the ratio $L \rho_{\rm hedg} /
\rho_{\rm discl}$ within this time range was of the order of
$10$--$100$. The case $\rho_{\rm discl}=10^6 \,{\rm cm}^{-2}$, 
$\rho_{\rm hedg}=10^8
\,{\rm cm}^{-3}$, chosen above to obtain the upper solid curve in
Fig.~\ref{FIG:Wong}, falls within this range.  Consequently, it cannot
be ruled out that the observed approach to $S(k) \propto k^{-5}$
from above in Fig.~7 of Ref.~\cite{Wong93} is due to the presence of
hedgehogs, even though if the relative proportion of populations of
hedgehogs and disclinations in the experiments of
Refs.~\cite{Wong92,Wong93} is similar to that found in the experiments
of Refs.~\cite{Chuang93} (which was performed on a different nematic material),
such a possibility should be considered unlikely.   A more likely possibility 
is that a portion of the length of the disclinations in the system 
is of the wedge type (see Sec.~\ref{SEC:RingsInUni}). 
Any curved configuration of a wedge disclination will lead to an 
order-parameter configuration in the vicinity of the disclination 
that resembles a part of a (radial or hyperbolic) 
hedgehog \cite{MZ+TCL}. (The special case of a disclination loop that is 
wedge-like everywhere along its length was discussed in 
Sec.~\ref{SEC:RingsInUni}; in this case, a full hedgehog configuration 
is obtained outside of the loop.) Although the twist configuration 
of a disclination line is energetically preferable to the 
wedge configuration for usual values of the nematic elastic 
constants \cite{Chandrasekhar}, wedge-type segments of disclination 
loops in a system undergoing phase ordering may be generated dynamically, 
and lead to a $k^{-6}$ contribution to the nematic structure factor. 

We complete this Section by a brief discussion of the dynamical scaling of
the structure factor.  Three different length scales --- the
disclination separation $L_{\rm discl}$, the hedgehog separation
$L_{\rm hedg}$, and the microscopic length scale $8 \,{\AA}$ associated with
thermal fluctuations --- appear in the large-$k$ form of the nematic
structure factor, Eq.~(\ref{combine}). Is this inconsistent with the
property of dynamical scaling of the structure factor, according to
which there exists a (time-dependent) length scale $L(t)$ such that
$k^3 S(k)$ depends only on the dimensionless scaling variable $y
\equiv k L(t)$?  Let us choose the disclination separation $L_{\rm discl}$
as the characteristic length $L(t)$.  From Eq.~(\ref{combine}), we
then have
\begin{equation}
k^3 S(y) = {3\pi^3} {1 \over y^2} + 
{36\pi^{4} \over y^3} \left[{L_{\rm discl}(t) \over L_{\rm hedg}(t)}\right]^3 
+ y {8 {\AA} \over L_{\rm discl}(t)} \,.
\label{scaling}
\end{equation}
Clearly, $k^3 S(k)$ does not in general satisfy the scaling form $f(k
L_{\rm discl}(t))$.  At asymptotically late times, however, 
$L_{\rm discl} \gg 8 {\AA}$ 
and, consequently, the term in Eq.~(\ref{scaling}) associated with
thermal fluctuations becomes negligible. Experimental evidence
\cite{Chuang93} indicates that at late times after the quench, the
hedgehog separation $L_{\rm hedg}(t)$ grows faster than the disclination
separation $L_{\rm discl}(t)$.  Consequently, the second term in
Eq.~(\ref{scaling}) (associated with hedgehogs) also becomes negligible,
asymptotically.  Our analysis is therefore consistent with the dynamical
scaling of the structure factor at asymptotically late times.  As we
saw earlier in this Section, however, the scattering contributions
from thermal fluctuations and from nematic hedgehogs can lead to
significant transient effects under typical experimental conditions.
\section{Biaxial nematic systems}
\label{SEC:BiaxNemSys}

Thus far, our discussion of defects in nematics has concentrated on
the case of {\it uniaxial\/} systems. It is straightforward to generalize this
discussion to the case of {\it biaxial\/} \cite{de Gennes} 
systems.  As there are no
topologically stable point defects in a three-dimensional biaxial
nematic \cite{Mermineg}, we need only consider line defects.  Biaxial
nematics admits four topologically distinct classes of line
defects~\cite{Mermineg}, which are disclination lines distinguished by
the angles of rotation of the uniaxial director ${\bf u}$ and the
biaxial director ${\bf b}$ [defined in Eq.~(\ref{paramQ})] around the
defect core.  In the $C_x$ class of disclinations ${\bf u}$ rotates by
$\pm 180^\circ$ and ${\bf b}$ does not rotate; in $C_y$ disclinations
${\bf u}$ does not rotate and ${\bf b}$ rotates by $\pm 180^\circ$; in
$C_z$ disclinations both ${\bf u}$ and ${\bf b}$ rotate by $\pm
180^\circ$; finally, in $\overline C_0$ disclinations either ${\bf u}$
or ${\bf b}$ (or both) rotate by $360^\circ$. These four distinct
disclination types were observed experimentally in a thermotropic
nematic polymer~\cite{De'Neve92}, and their properties where found to
be in agreement with the predictions of the topological classification
scheme.

In the minimum-energy configuration [with free energy given by 
Eq.~(\ref{oneFrank}] 
for the $C_x$, $C_y$, and $C_z$
defects, the $180^\circ$ rotations of the ${\bf u}$ and ${\bf b}$
director are uniform  \cite{Kobdaj93}.  The configuration of a $C_x$
disclination is correspondingly given (up to a global rotation) by
Eq.~(\ref{paramQ}) with
\begin{mathletters}
\begin{eqnarray}
{\bf u}({\bf x})
&=&
\big(
\cos{1\over{2}}\phi({\bf x}),
\sin{1\over{2}}\phi({\bf x}),
0\big), 
\\
{\bf b}({\bf x})
&=&
\big(0,0,1\big),
\\
{\bf v}({\bf x})
&=&
\big(
\cos\big({1\over{2}}\phi({\bf x})+{\pi\over{4}}\big),
\sin\big({1\over{2}}\phi({\bf x})+{\pi\over{4}}\big),
0\big),
\end{eqnarray}
\end{mathletters}%
where 
$\phi({\bf x})$ is the polar angle in the plane perpendicular to the
disclination line. The correlation function~(\ref{EQ:defcrnem}) for this
configuration can be expressed (up to an additive constant) as the
correlation function~(\ref{EQ:map}) for the uniaxial nematic
disclination, multiplied by the biaxiality-strength--dependent factor
$R(3 S_1 + S_2)^2/4$, where $R\equiv 3S_1^2/(3 S_1^2 + S_2^2)$ is the
ratio of the uniaxial ($S_2=0$) and biaxial ($S_2 \ne 0$) normalization
factors $M_{\rm nem}$.  The corresponding factors for the case of the
$C_y$ and $C_z$ disclinations are likewise readily evaluated.  By using
the result (\ref{EQ:finaldiscl}) for the uniaxial disclination, we
obtain the structure factors of the biaxial
disclinations $C_x$, $C_y$ and $C_z$:
\begin{mathletters}
\begin{eqnarray}
S(k)_{\rm C_x}
&=&
{9 \pi^2 \over 8} {L_x \over V}
{S_1^2 (3 S_1 + S_2)^2 \over 3 S_1^2 + S_2^2}  
{1 \over k^5}\,,
\label{finalbiax180X}
\\
S(k)_{\rm C_y} 
&=&
{9 \pi^2 \over 2} {L_y \over V}
{S_1^2 S_2^2 \over 3 S_1^2 + S_2^2}
{1 \over k^5} \,,
\label{finalbiax180Y}
\\
S(k)_{\rm C_z} 
&=&
{9 \pi^2 \over 8} {L_z \over V}
{S_1^2 (3 S_1-S_2)^2 \over 3 S_1^2 + S_2^2}
{1 \over k^5}\,.
\label{finalbiax180Z}
\end{eqnarray}
\end{mathletters}%
It 
should be noted that in the case $3 S_1=S_2$, the amplitude in the Porod
law for the $C_z$ defect is zero, whereas the amplitudes for the $C_x$
and $C_y$ defects are non-zero and equal to each other [see
Eqs.~(\ref{finalbiax180X})--(\ref{finalbiax180Z})].  This reflects the
fact that for $3S_1=S_2$, the order parameter 
given by Eq.~(\ref{paramQ}) 
describes
a {\it uniaxial discotic\/} phase \cite{discotic} with uniaxial axis
${\bf v} = {\bf u} \times {\bf b}$. Correspondingly, the $C_z$ 
configuration in this case does
not represent a defect (as ${\bf v}$ does not rotate), whereas the $C_x$
and $C_y$ configurations are equivalent.

It remains to consider the ${\overline C_0}$ disclination. For the case
$S_1>S_2$ (i.e., needle-like ordering \cite{discotic}), the minimum-energy 
configuration of type ${\overline C_0}$ is given by
Eq.~(\ref{paramQ}) with
\begin{mathletters}
\begin{eqnarray}
{\bf u}({\bf x})
&=&
\big(0,0,1\big),
\\
{\bf b}({\bf x})
&=&
\big(
\cos\phi({\bf x}),
\sin\phi({\bf x}),
0\big), 
\\
{\bf v}({\bf x})
&=&
\big(
\cos\big(\phi({\bf x})+\pi/4 \big),
\sin\big(\phi({\bf x})+\pi/4 \big),
0\big),
\end{eqnarray}
\end{mathletters}%
where 
$\phi({\bf x})$ is the polar angle in the plane perpendicular to the
disclination line. The correlation function~(\ref{EQ:defcrnem}) for the
point defect in the planar cross-section through this disclination can
then be expressed as
$R(2S_2/3S_1)^2$,  
By using Eq.~(\ref{stringres}) one obtains the structure factor per 
unit length of the ${\overline C_0}$ disclination: 
\begin{equation}
S(k)_{\rm \overline C_0}=
12\pi^2 {L_0 \over V}
{S_2^2\over 3S_1^2+S_2^2}
{1\over k^5}\,.
\label{finalbiax360a}
\end{equation}
For the case $S_1<S_2$ (i.e., discotic ordering~\cite{discotic}) 
the ${\bf v}$ director 
(and not the ${\bf u}$ director, in contrast with the needle-like case)
corresponds to the eigenvalue of the order-parameter tensor
$Q_{\alpha\beta}$ largest in absolute value, and the lowest energy 
configuration that can be taken
by the ${\overline C_0}$ disclination has directors given by
\begin{mathletters}
\begin{eqnarray}
{\bf u}({\bf x})
&=&
\big(
\cos\phi({\bf x}),
\sin\phi({\bf x}),
0\big), 
\\
{\bf b}({\bf x})
&=&
\big(
\cos\big(\phi({\bf x})+{\pi\over{4}}\big),
\sin\big(\phi({\bf x})+{\pi\over{4}}\big),
0\big),
\\
{\bf v}({\bf x})
&=&
\big(0,0,1\big)\,.
\end{eqnarray}The 
corresponding structure factor is 
\begin{equation}
S(k)_{\rm \overline C_0}
={8\over 3} {L_0 \over V} \pi^2{(3S_1^2-S_2)^2\over 3S_1^2+S_2^2} 
{1 \over k^5}\,.
\label{finalbiax360b}
\end{equation} 
\end{mathletters}%
In general, the large-$k$ structure factor of a biaxial nematic system 
is given by the sum of the expressions 
(\ref{finalbiax180X}-\ref{finalbiax180Y}) and (\ref{finalbiax360a}) or 
(\ref{finalbiax360b}).

Finally, we note that in a 
biaxial nematic film (i.e., a system having
two spatial dimensions, but the full $3 \times 3$ tensorial nematic 
parameter), $C_x$, $C_y$, $C_z$ and ${\overline C_0}$ are point-like
defects (the dynamics of which was studied in~\cite{2dnem}).  The
Porod-law amplitudes for these defects are obtained from the amplitudes
in the results~(\ref{finalbiax180X})--(\ref{finalbiax180Z}),
(\ref{finalbiax360a}) and (\ref{finalbiax360b}) by dividing by $\pi$
[see Eq.~(\ref{stringres})], and the corresponding Porod-law exponents take
the value 4 instead of 5.

\section{Corrections to Porod's law}
\label{SEC:CorrToPorod}

In this section we briefly discuss several effects---inter-defect
interactions, inequality among the three elastic constants in the
nematic free energy, curvature of linear defects, and the presence
of the defect core---that were not fully taken into account during our
calculations in the previous sections and may, under certain
circumstances, lead to modifications of our results.

An interesting question, already alluded to in
Sec.~\ref{SEC:GenSk}, 
is whether the
order-parameter configuration in the vicinity of the defect core is 
or is not affected by the interactions with the other defects present in
the system. [In order to avoid confusion with the core region, where
the order parameter magnitude is reduced, we shall refer to the region
close to, but outside, the core, as the \lq\lq central\rq\rq\ region 
of the defect.]
It is important to distinguish between static and dynamic
effects. In the case of the $O(2)$ vector model with the standard gradient free
energy, given by Eq.~(\ref{ONenergy}), the minimum-energy
configuration of a collection of defects with fixed locations is obtained
simply by taking the superposition of the angles characterizing the
order parameter ${\bf m}$ around isolated defects; their central
regions are therefore unaffected. A deformation does occur, however,
once the defects are allowed to move \cite{Pismen}; for sufficiently
slow defect velocities, this effect may be neglected. 

The situation is different in the $O(3)$ vector-model case. Here,
Ostlund \cite{Ostlund} showed that in the minimum-energy configuration
of a monopole-antimonopole pair separated by a fixed distance, the
gradient free energy is concentrated along a string connecting the two
point defects. We are not aware of any investigation of how this
result might be modified in a dynamical situation. Ostlund's result 
does suggest, however, that the order-parameter configuration of
monopoles and antimonopoles in a phase-ordering $O(3)$ vector system
in $d=3$ may substantially differ from the radially symmetric
configuration assumed during the calculation of the
corresponding Porod law in Sec.~\ref{SEC:PDinOnSymm} of the present Paper.

Consider, then, an arbitrary point defect 
 in the $O(N)$ vector model in $d=N$ dimensions.  
We denote the corresponding order-parameter configuration 
by ${\bf \Phi}({\bf r})$, where the radius-vector ${\bf r}$ 
originates at the core of the defect.
As
already indicated in Sec.~\ref{SEC:GenSk}, the value $\xi=2 d$ of
the Porod exponent for point defects 
is suggested by simple dimensional analysis,
independently of the exact form of ${\bf \Phi}({\bf r})$. This argument
can be made more precise if we consider configurations in which 
${\bf \Phi}({\bf r})$ depends only on the direction ${\bf {\hat r}}$, 
and not the magnitude $r$, of the radius-vector ${\bf r}$.
We may write 
${\bf \Phi}({\bf k}) 
\equiv \int d^dr\, e^{i {\bf k}\cdot{\bf r}} {\bf \Phi}({\bf r}) 
= \int d^dr\, e^{i k {\bf {\hat k}}\cdot{\bf r}} {\bf \Phi}({\bf r})$; 
by assumption, ${\bf \Phi}(k^{-1} {\bf r}) = {\bf \Phi}({\bf r})$, 
which, combined with the substitution ${\bf r} = k^{-1} {\bf y}$,  
yields  
${\bf \Phi}({\bf k}) 
= k^{-d} \int d^dy\, e^{i {\bf {\hat k}} \cdot{\bf y}} {\bf \Phi}({\bf y}) 
\equiv k^{-d}\,\, {\bf \Phi}({\bf {\hat k}})$.
 Consequently, the
structure factor Eq.~(\ref{EQ:wigner}) is of the form 
$S({\bf k}) = A({\bf {\hat k}}) k^{-2 d}$, where the Porod amplitude
$A({\bf {\hat k}}) 
\equiv M_{\rm O(N)}^{-1} {\bf \Phi}({\bf {\hat k}}) 
\cdot {\bf \Phi}(-{\bf {\hat k}}) $ 
now depends on the orientation of ${\bf k}$. The
Porod exponent $\xi= 2 d$, expressing how $S({\bf k})$ depends on the
{\it magnitude} of ${\bf k}$, is, however, unchanged compared to the
radially symmetric case.  As for the value of $A({\bf {\hat k}})$, we
may make the following argument using the identity relating the
standard form of the gradient free energy $E$ to the structure factor
$S({\bf k})$:
\begin{equation}
E \equiv {\kappa \over 2} 
\int d^d x\, \nabla_i \Phi_j ({\bf x}) \nabla_i \Phi_j({\bf x})
    =   {\kappa \over 2} \int\,d^dk \,k^2 S({\bf k}) = 
{\kappa \over 2} \big\langle A({\bf {\hat k}}) \big \rangle _{\rm ang}
\int^{\xi^{-1}}_{D^{-1}}\,d k \,k^{2 - 2 d} \,.
\label{esk}
\end{equation}
Here $\big\langle A({\bf {\hat k}}) \big \rangle _{\rm ang}$ denotes the
angular average of $A({\bf {\hat k}})$ over all orientations ${\bf {\hat k}}$,
 $\kappa$ is a positive
elasticity constant, $\xi$ is the core size of the defect and $D$ is
the large-distance cutoff necessary in the calculation of the free
energy. Equation~(\ref{esk}) implies that $\big\langle
A({\bf {\hat k}}) \big \rangle _{\rm ang}$ reaches its minimum for the
order-parameter configuration ${\bf\Phi}({\bf {\hat r}})$ that
minimizes the free energy $E$ \cite{alternative}. In the case of the $O(N)$
monopole with charge $+1$, the configuration minimizing $E$ is
precisely the radial configuration used by us in
Sec.~\ref{SEC:PDinOnSymm}, and we conclude that our result
Eq.~(\ref{EQ:MagRes}) provides a {\it lower bound\/} for the value of
the Porod amplitude in a macroscopically isotropic system \cite{Andrew}.

Note, however, that the property of 
$\big\langle A({\bf {\hat k}}) \big \rangle _{\rm ang}$ formulated above 
 applies equally
well for defects that do not possess radial symmetry in the minimum-energy
configuration, such as monopoles with higher-integer charges in the
$O(N)$ vector model, or strength-$1/2$ disclinations in nematic liquid
crystals [in the latter case, $E$ in Eq.~(\ref{esk}) is replaced by
the Frank free energy in the one-constant approximation,
Eq.~(\ref{oneFrank})]. We thus reach the conclusion that in nematics
for which the three elastic constants (see Sec.~\ref{SEC:intronem}) are
not equal to each other and consequently, the configuration of the
nematic director around a disclination differs from the configuration
minimizing Eq.~(\ref{oneFrank}), the angle-averaged structure factor
in a phase-ordering system will have Porod amplitudes larger than
those derived by us in Secs.~\ref{SEC:hedge}--\ref{SEC:exper}, even when 
the central regions of the defects are {\it not\/} deformed by inter-defect 
interactions. 

While deriving the results for linear defects [$O(3)$ vortex lines in
Sec.~\ref{SEC:VortexLines} and nematic disclinations in
Secs.~\ref{SEC:DisInUni} and \ref{SEC:RingsInUni}], we assumed that
the curvature $1/R$ of the defect lines was negligible (i.e. $R k \gg 1$).  
In general, we can expect corrections to the basic Porod law $k^{-5}$ 
that are suppressed by factors of $R k$, thus resulting in contributions 
to $S(k)$ decaying as higher power laws ($k^{-\eta}$ with $\eta > 5$). 
The case of curved wedge-type disclinations (discussed near the end 
of Sec.~\ref{SEC:DisInUni}), for which $\eta=6$, serves as an example.

Similarly, we can expect higher-power terms to arise from the
presence of the core region of the defects. 
As the crossing of the boundary between the core region and the
outside region of a defect is associated with a change in the
magnitude of the order parameter, we may obtain an upper-bound estimate of
the corresponding contribution $S_{\rm core}(k)$ 
to the structure factor by regarding
the core boundary as a domain wall in a scalar system, leading (in
a 3-dimensional system) to $S_{\rm core}(k) \simeq a k^{-4}$, where $a$ is
the domain wall area. For a monopole $a \simeq \xi^2$, where $\xi$ is
the core size, and the ratio of $S_{\rm core}(k)$ to the standard monopole
contribution $k^{-6}$ is of the order of $(\xi k)^{-2}$.  For a
string defect of length $L$, we have $a \simeq \xi L$, and the ratio of
$S_{\rm core}(k)$ to $L k^{-5}$ is of the order of $\xi k$. 
For typical core sizes $\xi$ of the order of 
$10 \AA$ and for scattering wave-vector values $k \leq 10^{-4}
\AA$ accessible with visible light, we have $\xi k \geq 1000$, and the
contributions to $S(k)$ from the defect core are expected to be 
negligible for both
monopoles and strings.

\section{Conclusions}
\label{SEC:Concl}

In this Paper we have presented a detailed discussion 
of the influence of topological defects on the large-wavevector 
behavior of the 
structure factor $S({\bf k})$ in nematic liquid crystals. 
The presence of topological defects leads to power-law contributions to 
$S({\bf k})$
of the Porod form $\rho A k^{-\xi}$, where $\rho$ is the 
number density of a given
type of defect, $A$ is a dimensionless amplitude, 
and $\xi$ is an integer-valued exponent. 
We have computed the values of the Porod exponents and amplitudes  
for the various types of topological defects present 
in uniaxial and biaxial nematics, 
and have discussed the competition between contributions to $S({\bf k})$ due 
to defects and to thermal fluctuations.

Our main results are summarized in 
Tables~\ref{table_2d} (for nematic films) and
\ref{table_3d} (for bulk nematic systems). To obtain the
short-distance structure factor of a 
nematic system containing
topological defects, the expressions given in
Tables~\ref{table_2d} or \ref{table_3d} should be multiplied by the number 
densities of the corresponding defects and then added.  Here, the defect
number density is defined as the number of point defects, resp. the
total length of line defects, per unit area (in a two-dimensional system) 
or unit
volume (in a three-dimensional system).  
In addition, the total structure factor
contains a power-law contribution due to transverse thermal
fluctuations of the nematic director, this contribution 
being given (in a bulk system)
approximately by $8 {\AA} / k^2$.  
The resulting form of the structure factor 
$S(k)$ is valid for $k$ ranging from the inverse typical 
separation of defects to the inverse defect-core size.

\begin{table}[b]
\caption{Power-law contributions to the structure factor arising from 
topological defects in systems 
with spatial dimension $d=2$. Results are given per unit defect density 
(see Sec.~\protect\ref{SEC:Concl}). For biaxial nematics, $S_1$ and 
$S_2$ denote the uniaxial and biaxial amplitudes of the nematic order 
parameter, and $C_x$, $C_y$ and $C_z$ denote the three topologically 
distinct $180^{\circ}$ defect types (see Sec.~\protect\ref{SEC:BiaxNemSys}).}
\begin{tabular} {|c|c|l|c|c|}
& Uniaxial & Biaxial & O(2) vector & O(3) vector \\
& nematic & nematic & model & model \\
\tableline
\hline
$180^{\circ}$ disclination point & $3\pi^2 k^{-4}$ 
& $C_x:\,\,\,\,\,{9 \pi^2 \over 8}{S_1^2 (3 S_1 + S_2)^2 \over 3 S_1^2 + S_2^2}  k^{-4}$
& does not exist & topologically\\ 
&&$C_y:\,\,\,\,\,{9 \pi^2 \over 2}{S_1^2 S_2^2 \over 3 S_1^2 + S_2^2} k^{-4}$
&&unstable\\ 
&&$C_z:\,\,\,\,\,{9 \pi^2 \over 8}{S_1^2 (3 S_1-S_2)^2 \over 3 S_1^2 + S_2^2} k^{-4}$
&&\\ \hline
$360^{\circ}$ disclination point & topologically & 
$(S_1>S_2):\,\,\,\,\,{S_2^2\over 3S_1^2+S_2^2}k^{-4}$ & $4 \pi^2 k^{-4}$ 
&topologically \\
&unstable&$(S_1<S_2):\,\,\,\,\,{8\over 3}\pi^2{(3S_1^2-S_2)^2\over 3S_1^2+S_2^2} k^{-4}$
&&unstable\\ 
\end{tabular}
\label{table_2d}
\end{table}

\begin{table}[hbt]
\caption{Power-law contributions to the structure factor arising from 
topological defects in systems 
with spatial dimension $d=3$. The results given in the first three rows 
are given per unit defect density 
(see Sec.~\protect\ref{SEC:Concl}).
The results in the fourth (fifth) row correspond to 
a single twist (wedge) circular disclination loop of radius $R$ 
in a system of volume $V$;
$\zeta$ denotes the angle between the scattering wave-vector ${\bf k}$ 
and the loop axis.
For biaxial nematics, $S_1$ and 
$S_2$ denote the uniaxial and biaxial amplitudes of the nematic order 
parameter, and $C_x$, $C_y$ and $C_z$ denote the three topologically 
distinct $180^{\circ}$ defect types (see Sec.~\protect\ref{SEC:BiaxNemSys}).}
\begin{tabular} {|c|c|l|l|l|}
& Uniaxial & Biaxial & O(2) vector & O(3) vector \\
& nematic & nematic & model & model \\
\tableline
\hline
& & & & \\

Hedgehog defect& $36\pi^{4} k^{-6}$ & does not exist
& does not exist& $12 \pi^{3} k^{-6}$\\ \hline
 
&& $C_x:\,\,\,\,\,{9 \pi^3 \over 8}{S_1^2 (3 S_1 + S_2)^2 \over 3 S_1^2 + S_2^2}   k^{-5}$
&&\\

$180^{\circ}$ disclination line &${3\pi^3}  k^{-5}$
& $C_y:\,\,\,\,\,{9 \pi^3 \over 2}{S_1^2 S_2^2 \over 3 S_1^2 + S_2^2}  k^{-5}$
&does not exist&does not exist\\ 

(orientational average)&&$C_z:\,\,\,\,\,{9 \pi^3 \over 8}{S_1^2 (3 S_1-S_2)^2 \over 3 S_1^2 + S_2^2}  k^{-5}$
&&\\ \hline

$360^{\circ}$ disclination line & topologically
& $(S_1>S_2):\,\,\,\,\,\pi {S_2^2\over 3S_1^2+S_2^2} k^{-5}$
& $4 \pi^3  k^{-5}$ & topologically\\ 

(orientational average)&unstable&$(S_1<S_2):\,\,\,\,\,{8\over 3}\pi^3{(3S_1^2-S_2)^2\over 3S_1^2+S_2^2}  k^{-5}$
&&unstable\\ \hline

&& $C_x:\,\,\,\,\,{9 \pi^3 \over 2}{S_1^2 (3 S_1 + S_2)^2 \over 3 S_1^2 + S_2^2}  
{R \over V} k^{-5} /|\sin{\zeta}|$
& & \\

Twist disclin.~loop& $12 \pi^3 {R \over V} k^{-5} /|\sin{\zeta}|$&
$C_y:\,\,\,\,\,{18 \pi^3}{S_1^2 S_2^2 \over 3 S_1^2 + S_2^2} {R \over V}
 k^{-5} /|\sin{\zeta}|$
& $16 \pi^3 {R \over V} k^{-5} /|\sin{\zeta}|$ & does not exist \\ 

&&$C_z:\,\,\,\,\,{9 \pi^3 \over 2}{S_1^2 (3 S_1-S_2)^2 \over 3 S_1^2 + S_2^2} 
{R \over V} k^{-5} /|\sin{\zeta}|$
&&\\ \hline

Wedge disclin.~loop& $\cases{ {36 \pi^4 \over V}k^{-6},& 
$k \leq R^{-1}$\cr
             12 \pi^3 {R \over V} k^{-5}/|\sin{\zeta}|,& $k > R^{-1}$
                                              }$ & does not exist
& does not exist & does not exist\\ 
\end{tabular}
\label{table_3d}
\end{table}

To avoid confusion, we remind the reader 
that our results were calculated for a structure factor $S({\bf k})$
defined so that the real-space correlation function $C({\bf r})$ is
normalized to unity at $r=0$ (and, consequently, our structure factor has the
dimension of volume). Thus, our results differ from the un-normalized
structure factor (with dimension of volume squared) by the
normalization factor $M_{\rm norm}=(V / 2) (3 S_1^2 + S_2^2)$ (where
$V$ is the system volume or area, $S_1$ is the (uniaxial) order-parameter 
magnitude, and $S_2$ is the strength of biaxial ordering).
Recall, in addition, that the scattered-light intensity is directly
proportional 
to the structure factor $S({\bf k})$ only in the case of unpolarized 
light scattering. Finally, recall (Sec.~\ref{SEC:CorrToPorod}) that the 
values of the Porod amplitudes 
(but not those of the Porod exponents) in the general case of 
unequal 
elastic constants in the Frank free energy (\ref{genFrank}) are expected 
to differ from the values (derived in the one-constant approximation) 
listed in Tables \ref{table_2d} and \ref{table_3d} .

For comparison, we also show in Tables \ref{table_2d} and \ref{table_3d} 
the results for the corresponding defects (when they exist) in 
$O(2)$ and $O(3)$ symmetric vector-model systems. In these systems, 
the normalization factor $M_{\rm norm}$ is given by $V s^2$, 
where $s$ is the order-parameter magnitude. Identical results for some 
of the $O(N)$ defect configurations were previously obtained in 
Ref.~\cite{Bray-amplitudes}. 

We conclude this Paper by highlighting some features of our results. 
As discussed in Sec.~\ref{SEC:GenSk}, the value of the 
Porod exponent $\xi$ 
can be correctly anticipated purely on dimensional grounds. 
It should be noted, 
however, that the value of the dimensionless Porod amplitude $A$ 
often differs considerably from unity. 
For example, in the case of the nematic hedgehog 
defect, we obtained $A=36 \pi^4 \simeq 3500$. In addition, the Porod 
amplitude depends significantly on the order parameter in question 
(and not just on the spatial and defect dimensionalities). 
For example, the Porod amplitude for 
the nematic hedgehog exceeds the amplitude for the $O(3)$ monopole in $d=3$ 
by a factor of $3 \pi$.

In Secs.~\ref{SEC:thermal} and \ref{SEC:exper}, we analyzed in detail 
the conditions under which 
the Porod tail of the structure factor 
is not overshadowed by the power-law contribution from transverse 
thermal fluctuations of the nematic director. 
[Similar considerations apply to any system possessing continuous symmetry 
of the order parameter.]
We concluded
that for experimentally accessible defect densities in a nematic 
phase-ordering 
experiment, 
the Porod tail should be 
observable over a range of 1 to 3 decades in $k$; the range 
of observability is limited by defect interactions at small $k$ 
and by thermal fluctuations at large $k$.  
In the experiments reported in Refs.~\cite{Wong92,Wong93}, the measured 
structure factor did 
exhibit a Porod tail over at least 1 decade in $k$, but the crossover at 
high $k$ to the fluctuation-dominated regime was not observed. 
This was found to be consistent with our theoretical predictions 
under the conditions of these experiments. 

The crossover from $S(k) \propto k^{-5}$ (in a system 
where disclinations dominate over hedgehogs) to $S(k) \propto k^{-2}$ 
(the thermal-fluctuation-dominated regime) was predicted 
in Sec.~\ref{SEC:exper} 
to occur approximately at the scattering wave-vector value 
$k_{\rm u} = (3 \pi^3 / 8 {\AA})^{1/3} L_{\rm discl}^{-2/3}$, where 
$L_{\rm discl}$ is the total disclination length per unit volume of 
the system. 
As this crossover is very sharp, and is not affected by defect interactions, 
measuring $k_{\rm u}$ can give a precise estimate of the total disclination 
length. We are not aware of any use of this method in the experimental 
literature to date. 
The proportionality of the Porod tail to the defect density has been 
used previously \cite{Wong92,Wong93} to extract the power law characterizing 
the decay of disclination length; extracting the absolute defect density 
in this way, however, would require the knowledge of the (rarely 
experimentally available) 
normalized structure factor. The location of the crossover 
$k_{\rm u}$, on the other hand, can be extracted from the unnormalized 
structure factor.  

In Sec.~\ref{SEC:RingsInUni}, we contrasted the contributions 
to the structure factor 
arising from twist-type and wedge-type disclination loops. 
The wedge-type loop of radius $R$ gives a $k^{-6}$ contribution 
at $k<R^{-1}$ in addition to the usual $k^{-5}$ disclination contribution 
at $k>R^{-1}$, as it has the structure of a nematic 
hedgehog at large distances 
from the loop. Likewise, any curved disclination segment of the wedge type 
gives rise to a $k^{-6}$ contribution. 
In Sec.~\ref{SEC:exper}, we attempted to connect these findings 
to the observation 
that the structure factor in phase-ordering uniaxial nematics 
approaches the $k^{-5}$ power law 
at high $k$ through effective exponents larger 
than 5 (and close to 6) at intermediate $k$. Such a behavior of the structure 
factor had been found to occur in experiments \cite{Wong92,Wong93} as well as 
in numerical simulations~\cite{Blundell94}, but not in approximate analytical 
theories \cite{Bray-nematic}. We concluded that it was, in principle, 
possible, 
but nevertheless unlikely, that the ``approach from above'' to the exponent 
$\chi=5$ observed in the experiments was due to $k^{-6}$ contributions from 
hedgehog defects. It remains a challenge for future work to determine 
whether this behavior of the structure factor is 
caused by hedgehog defects, curved wedge-type disclination segments, or 
correlations 
that fall beyond the range of the Porod regime.    

\acknowledgements
M.~Z. wishes to thank A.~J.~Bray, T.~C.~Lubensky, A.~D.~Rutenberg, 
and B.~Yurke
for useful discussions.  This work was supported by the U.~S.~National Science
Foundation through Grants 
 DMR95-07366 (M.~Z.)  and 
DMR94-24511 (P.~M.~G.).




\begin{references}
\bibitem{twodimthermal} We do not discuss thermal fluctuations
at the lower criticial dimension $d=2$, where the situation
is more complicated.
\bibitem{Nigel}
This method was already used in Ref.~\protect\cite{Mondello92} to obtain
the Porod exponent for defects in the two-dimensional $O(2)$ symmetric 
vector model.
\bibitem{angle_note} If ${\bf k}$ is perpendicular to the plane of the loop, 
then from 
Eq.~(\ref{string}) we see that the total structure factor is 
$S({\bf k}) = 4 \pi R V^{-1} A^{(2)} k^{-4} \delta(k)$, where 
for $\delta(k)$ finite $R$  
has the meaning of a function with a sharp peak around $k=0$ 
having a half-width of the order of $\pi/R$. Consequently, we obtain 
a zero contribution 
to $S({\bf k})$ for $k>\pi/R$.
\bibitem{textures} In addition to the two types of singular defects (hedgehogs 
and disclinations), non-singular topologically stable objects---
topological textures \protect\cite{Mermin}---can occur in uniaxial
nematics. A single-texture configuration, however, does not result in 
a power-law contribution to the structure factor.
Furthermore, these objects were not found to occur
naturally in the phase-ordering experiments reported in Ref.~\cite{Chuang91}.
We shall not be concerned with topological textures in the present Paper.
\bibitem{field} We note, however, that if the sample is placed 
in an external
magnetic or electric field, large-$k$ thermal fluctuations are suppressed 
in the regions of the sample where the director is aligned along the field
  \protect\cite{de Gennes}. This can lead to an increase of the 
upper limit $k_u$
of the Porod regime.
\bibitem{alternative} It is worth noting that Eq.~(\protect\ref{esk}) 
can be used 
as an indirect means of evaluating the orientation-averaged 
Porod amplitude in the cases where 
the energy $E$ of the defect configuration is known.
\bibitem{wedge} 
In general, the director rotates in any way such that 
two antipodal points in the nematic order-parameter space $RP_2$ are connected
upon encircling the disclination core, and the disclination is not necessarily
of either the wedge type or the twist type.
\end{references}
\end{document}